\documentclass[journal]{IEEEtran}
\renewcommand{\baselinestretch}{0.98}
\usepackage{amsfonts}
\usepackage{amssymb}
\usepackage{amsthm}
\usepackage{amsmath,amsfonts,amssymb}
\usepackage[dvips]{graphicx}
\usepackage{subfigure}
\usepackage{verbatim}
\usepackage{setspace}
\usepackage{bm}
\usepackage{algorithmic} 
\usepackage[ruled,vlined]{algorithm2e}
\usepackage{cite}

\usepackage{changepage}
\usepackage{pdfpages}
\usepackage{color}
\usepackage{booktabs}
\usepackage{multirow}
\usepackage{makecell}

\allowdisplaybreaks

\IEEEoverridecommandlockouts

\begin{document}
\title{Movable-Antenna Relay Enhanced Multiuser Communications: Reinforcement Learning-Based Antenna Positioning and Power Allocation}
\author{\IEEEauthorblockN{Yuanjun Zhang, Lipeng Zhu, \textit{Senior Member, IEEE}, Haobin Mao, \textit{Graduate Student Member, IEEE}, Zhe Li, \textit{Graduate Student Member, IEEE}, and Zhenyu Xiao, \textit{Senior Member, IEEE}}
	
\thanks{
	Part of this work has been presented at the 2025 IEEE International Conference on Communications Workshops (ICC Workshops), Montreal, Canada~\cite{me2025MA}.
	
	Yuanjun Zhang, Haobin Mao, Zhe Li, and Zhenyu Xiao are
	with the State Key Laboratory of CNS/ATM, and the School of
	Electronics and Information Engineering, Beihang University 100191, China (e-mail:
	yuanjunzhang@buaa.edu.cn; maohaobin@buaa.edu.cn; zheli@buaa.edu.cn; xiaozy@buaa.edu.cn).
	
	Lipeng Zhu is with the State Key Laboratory of CNS/ATM and the School
	of Interdisciplinary Science, Beijing Institute of Technology, Beijing 100081,
	China (e-mail: zhulp@bit.edu.cn).
}
\vspace{-0.58 cm}
	
}

\maketitle


\begin{abstract}
	Movable antenna (MA) has emerged as a promising technology for enhancing wireless communication performance by flexibly reconfiguring wireless channels via antenna movement. In this paper, we propose an MA relay-enhanced multiuser communication system that adopts a half-duplex decode-and-forward (DF) relay scheme. The relay node is equipped with an MA array, and a buffer is employed to enhance the end-to-end transmission performance. Specifically, we formulate an optimization problem to maximize the minimum average achievable rate among multiple users over a given transmission period by dynamically optimizing the positions of MAs, transmit power, and beamforming matrices, subject to the constraints on antenna movement, inter-antenna spacing, and total energy budget for antenna movement and transmission. To address this non-convex problem with highly coupled variables, we propose a reinforcement learning (RL)-based algorithm, where a twin delayed deep deterministic policy gradient (TD3) algorithm is utilized to optimize the antenna positions and power allocation, while the transmit and receive beamforming solutions are obtained through closed-form expressions in each time slot. Numerical results validate the efficacy of the proposed RL-based algorithm and demonstrate the superiority of the proposed MA relay-enhanced multiuser communication system over the conventional fixed-position antenna (FPA)-based relay system and other MA-based benchmarks.
\end{abstract}


%
\IEEEpeerreviewmaketitle

\section{Introduction}
\IEEEPARstart{W}{ith} the rapid advancement of emerging wireless applications, such as autonomous driving and low-altitude networks, mobile communication systems are expected to meet increasingly stringent requirements for transmission reliability and spectral efficiency~\cite{saad20206G}. Against this backdrop, multiple-input multiple-output (MIMO) technology has been widely recognized as a key enabling technology and extensively investigated in both academia and industry~\cite{mimo2004,massive2014,massive2019}. Moreover, the migration of wireless communications toward higher-frequency bands, including millimeter-wave and terahertz bands, enables the deployment of extremely-large antenna arrays, thereby triggering the evolution from conventional MIMO systems toward massive MIMO and extremely large-scale MIMO (XL-MIMO) architectures~\cite{rapp2019Thz,rusek2013XLMIMO}. By employing multiple antennas at transceivers, MIMO systems are capable of exploiting the spatial degrees of freedom (DoFs) to achieve spatial multiplexing, diversity, and beamforming gains, thereby supporting high-rate and reliable transmissions~\cite{multiuser2004,mimo2009}. Despite these significant advantages, scaling up MIMO systems introduces an increasing number of antennas and radio frequency (RF) chains, resulting in prohibitive hardware costs and substantial signal processing overhead. In addition, the fixed antenna deployment at discrete positions in such MIMO systems fails to fully exploit the continuous spatial variations of wireless channels, which may limit the communication performance~\cite{Bj2016mimo}.

To address the inherent limitations of the conventional fixed-position antenna (FPA)-based MIMO systems, movable antenna (MA)~\cite{MA2024,MA2026}, also known as fluid antenna~\cite{zhu2024FAS}, has emerged as a promising technology to enhance wireless communication performance. Specifically, by flexibly adjusting the positions of antennas within predefined regions, MA-aided wireless communication systems offer enhanced spatial diversity and multiplexing performance with the same number of or even fewer antennas compared to conventional FPA-based systems~\cite{zhu2026,xiu2026MA,amhaz2026CoMP,liu2026NFC,chen2026MA}. In addition, the joint optimization of antenna positions and beamforming vectors provides additional beamforming flexibility, enabling enhanced signal transmission in desired directions and more effective interference suppression in undesired directions~\cite{zhu2026ToMA,ma2026ISAC,ke2026MA,yang2026MA}. Furthermore, the extremely large-scale MA system is capable of proactively establishing line-of-sight (LoS) channels between transceivers and reducing their path loss to further boost communication performance~\cite{fu2025MA,wang2025CFN}. In parallel, several other spatially reconfigurable antenna technologies, such as pinching antennas~\cite{ding2025PA,ding2026PA,xu2025PA} and rotatable antennas~\cite{zheng2026RA,zheng2026RACS,zhou2025RA}, have been proposed by exploiting either the position or orientation DoF of antennas for enhancing the communication performance of wireless communication systems.

Motivated by the substantial performance advantages offered by MAs, MA-aided wireless communication systems have attracted considerable research attention.
In~\cite{MA2024zhu}, an MA-aided multiuser uplink communication system was considered, where the users’ total transmit power was minimized by optimizing the MA positions at the users and the receive combining matrix at the base station (BS). The authors in~\cite{li2026MA}  investigated the MA-aided non-orthogonal multiple access (NOMA) systems, where the minimum achievable rate among all users was maximized by jointly optimizing the MA positions at the users, transmit beamforming, and the successive interference cancellation (SIC) decoding order through a two-loop iterative optimization algorithm.
Existing studies have also investigated the deployment of MAs at the base station (BS).
The authors in~\cite{wu2025move} investigated an MA-enabled multiuser multiple-input single-output (MISO) communication system, where a generalized Benders decomposition algorithm was proposed to obtain the globally optimal MA positions at the BS and beamforming vectors, and the total transmit power of the BS was minimized. The authors in~\cite{pi2024mu} proposed to deploy multiple MAs at the BS to receive uplink signals from multiple users, where the minimum achievable rate was maximized by jointly optimizing the MA positions, receive combining matrix, and users’ transmit power via a particle swarm optimization (PSO)-based iterative algorithm. In addition, the authors in~\cite{zhang2025mu} proposed an MA-aided multiuser hybrid beamforming scheme, where the sum rate was maximized by jointly optimizing the positions of movable sub-arrays and digital and analog beamforming through an alternating optimization (AO) framework.
In~\cite{mao2025MA}, the authors investigated the application of MAs in multiuser covert communications, where multiple MAs were deployed at the access point, and the covert sum rate was maximized by jointly optimizing the transmit beamforming and MA positions via a block successive upper-bound minimization-based algorithm.

It is noted that accurate channel state information (CSI) is essential for MA position optimization in the aforementioned systems. In this context, substantial research efforts have been devoted to CSI acquisition in MA-enabled communication systems~\cite{cao2024MA,cao2025MA,ma2023MA}.
In~\cite{cao2024MA}, the angles of departure (AoDs), angles of arrival (AoAs), and complex coefficients of the multipath components were jointly estimated through a compressed sensing-based method.
In~\cite{cao2025MA}, the channel estimation framework was extended from narrowband to wideband MA systems, where the sparse channel characteristics were reconstructed using a simultaneous orthogonal matching pursuit (SOMP)-based algorithm.
To reconstruct wireless channels throughout the movable region, the authors in~\cite{ma2023MA} proposed a successive transmitter-receiver compressed sensing (STRCS)-based channel reconstruction method to estimate the field-response information.

In relay systems, the limitations of conventional FPA-based architectures become more pronounced, as severe fading or interference over either the source-relay or relay-destination link can substantially degrade the end-to-end system performance.
In this context, the authors in~\cite{han2025re} investigated an MA relay-enhanced two-way communication system, where the minimum achievable rate was maximized by optimizing the MA positions via PSO, while maximal-ratio combining (MRC) was adopted for beamforming.
In~\cite{li2025re}, MA-enhanced amplify-and-forward (AF) and decode-and-forward (DF) relaying architectures were investigated. By jointly optimizing the MA positions and beamforming, the achievable end-to-end rate surpassed that of conventional FPA- and antenna selection-based schemes.
In~\cite{zhou2025re}, an MA-aided unmanned aerial vehicle (UAV) relay communication system was considered, where the relay rate was maximized by jointly optimizing the beamforming weights, antenna positions, UAV trajectory, and uplink-downlink time allocation.
In~\cite{li2024re}, a two-stage MA-aided AF relay system was investigated, where an AO algorithm integrating semidefinite programming (SDP), gradient ascent, and genetic algorithm (GA) was proposed to maximize the end-to-end signal-to-noise ratio (SNR).

\begin{figure*}[t]
	\vspace{0.2cm}
	\begin{center}
		\includegraphics[width=14 cm]{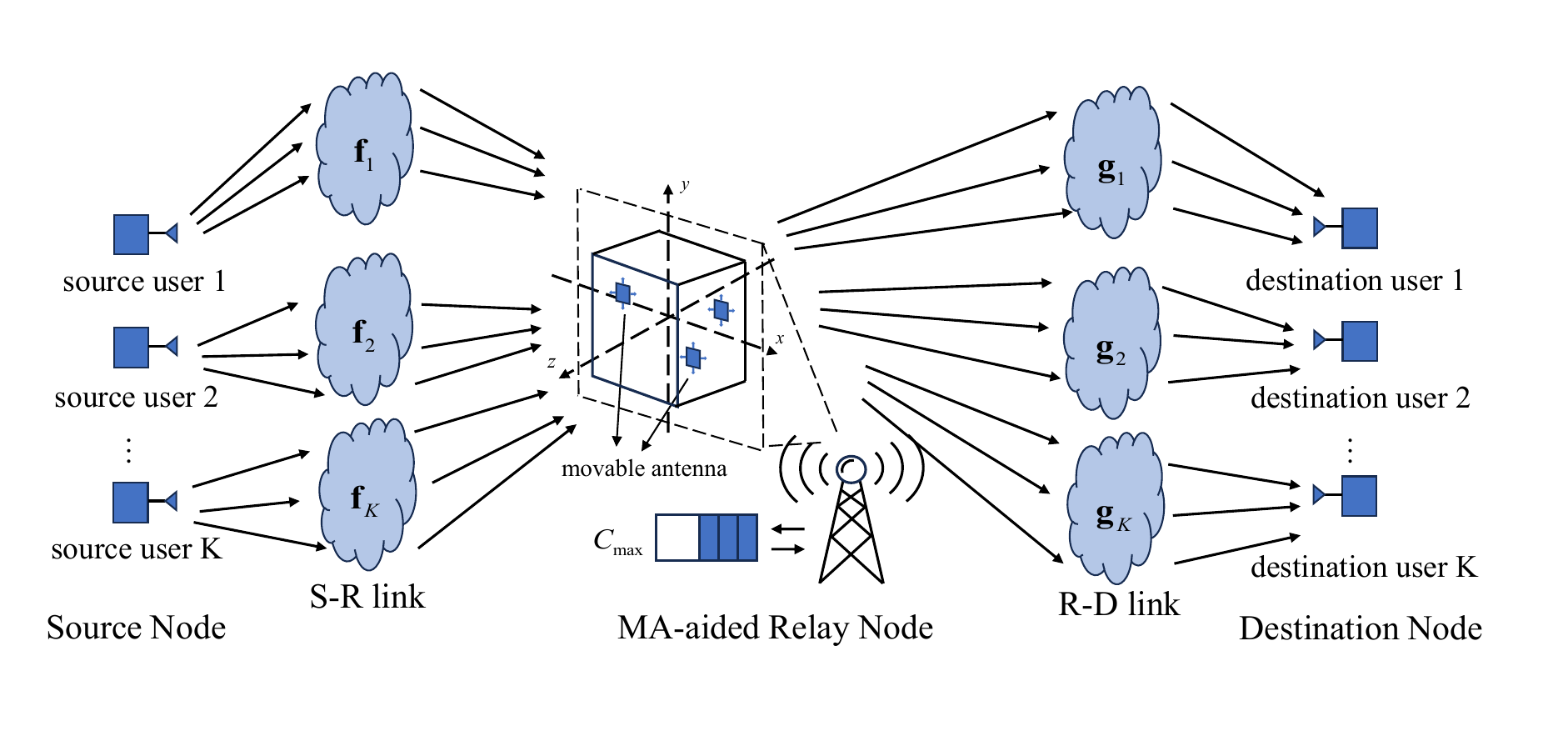}
		\caption{Illustration of the considered MA relay-enhanced multiuser system.}
		\label{fig:system}
	\end{center}
\end{figure*}

However, it is worth noting that existing studies on MA relay-enhanced communications have primarily focused on systems with a single transceiver pair. In practical relay systems with multiple communication pairs, the overall system performance can be significantly degraded by multi-pair transceiver interference, which cannot be effectively addressed by the existing solutions in the aforementioned works. To address this issue, we investigate an MA relay-enhanced multiuser communication system, where the spatial reconfigurability of MAs is exploited to mitigate multi-transceiver interference and improve the end-to-end communication performance. The main contributions of this paper are summarized as follows:

\begin{itemize}
	\item We consider a two-link MA relay-enhanced multiuser system, where the relay node is equipped with MAs to enable end-to-end half-duplex DF transmission. Moreover, a buffer is introduced to temporarily store the received data and exploit the capacity variations across the two links. Then, we design a relay operation protocol, where each time slot is divided into three stages to coordinate antenna movement, uplink transmission, and downlink transmission. To enhance the end-to-end communication performance for multiple user pairs, we formulate an optimization problem to maximize the minimum average end-to-end achievable rate among all user pairs by jointly optimizing the MA positions, power allocation, and beamforming matrices.
	\item To solve the formulated non-convex problem, we develop a DRL-based two-layer algorithm to address the resulting sequential decision-making problem. Specifically, the outer layer employs the twin delayed deep deterministic policy gradient (TD3) algorithm to optimize the MA positions and power allocation. The inner layer derives the corresponding beamforming matrices in closed form for each time slot.
	\item Extensive simulations are conducted to evaluate the performance of the proposed algorithm. Numerical results demonstrate that the proposed MA relay-enhanced system achieves a superior average achievable rate to conventional FPA-based systems. The results also verify the effectiveness of the buffer-aided relay scheme in improving the communication performance of the MA relay system.
\end{itemize}

The rest of this paper is organized as follows. In Section \uppercase\expandafter{\romannumeral 2}, we present the considered MA relay-enhanced multiuser communication system model and formulate the optimization problem. In Section \uppercase\expandafter{\romannumeral 3}, we develop the optimization algorithm based on DRL. Numerical results are provided in Section \uppercase\expandafter{\romannumeral 4}. Finally, we conclude this paper in Section \uppercase\expandafter{\romannumeral 5}. 

\textit{Notations}: $a$, $\mathbf{a}$, and $\mathbf{A}$ denote a scalar, a vector, and a matrix, respectively. $(\cdot)^{\mathrm{T}}$ and $(\cdot)^{\mathrm{H}}$ denote transpose and Hermitian transpose, respectively. $\mathcal{N}(0,\sigma^2)$ denotes the real Gaussian distribution with mean zero and variance $\sigma^2$. $\mathcal{CN}(0,\delta^2)$ denotes the circularly symmetric complex Gaussian (CSCG) distribution with mean zero and variance $\delta^2$. $\mathbb{C}$ represents the set of complex numbers. $\left|\cdot\right|$ denotes the amplitude of a complex number or the absolute value of a real number. $\lVert \cdot \rVert_2$ denotes the 2-norm of a vector. $\mathrm{diag}\{\mathbf{a}\}$ is a diagonal matrix with the element in row $i$ and column $i$ equal to the $i$-th element of vector $\mathbf{a}$. $\mathbf{I}_N$ denotes the $N \times N$ identity matrix.

\section{System Model and Problem Formulation}
As shown in Fig. \ref{fig:system}, we consider a two-link relay communication system, which consists of a shared relay node (R), multiple source nodes (S), and multiple destination nodes (D). The source and destination nodes consist of $K$ pairs of users, and each user pair can only exchange information through the common relay node for end-to-end transmission. Both the source and destination users are equipped with a single FPA, while $M$ MAs are deployed at the relay node. The relay communication adopts a DF half-duplex transmission scheme, where the communication process is divided into a multi-access communication (MAC) stage and a broadcast communication (BC) stage. Specifically, the relay node receives and decodes the signals during the MAC stage, and re-encodes them for transmission in the BC stage. In order to store the received information temporarily, we assume that a buffer at the relay node is shared by $K$ pairs of users, with a capacity of $C_{\mathrm{max}}$.

In this paper, we discretize the whole period $T$ into $N$ time slots to simplify the optimization of the MAs’ continuous trajectories. As $N$ becomes sufficiently large, the discrete antenna positions at multiple time slots can approximate the continuous trajectory. Then, the interval of each time slot $\Delta t = T/N$ is divided into three subslots for antenna movement, MAC, and BC. The time durations of three subslots are denoted by $\Delta t_{m}$, $\Delta t_{\mathrm{MAC}}$, and $\Delta t_{\mathrm{BC}}$, respectively.
In the $n$-th time slot, the local 3D coordinate of the $m$-th MA is denoted as $\mathbf{q}_m[n] = [x_m[n],y_m[n],z_m[n]]^{\mathrm{T}} \in \mathcal{C}_{\mathrm{r}}$, and the antenna position vector (APV) for all MAs is denoted as $\tilde{\mathbf{q}}[n]=\left[\mathbf{q}_1[n]^{\mathrm{T}},\mathbf{q}_2[n]^{\mathrm{T}},\dots,\mathbf{q}_M[n]^{\mathrm{T}}\right]^{\mathrm{T}}$. Without loss of generality, we assume that the moving region for MAs is cuboid, i.e., $\mathcal{C}_{\mathrm{r}} = [x_{\mathrm{r}}^{\mathrm{min}},x_{\mathrm{r}}^{\mathrm{max}}]\times[y_{\mathrm{r}}^{\mathrm{min}},y_{\mathrm{r}}^{\mathrm{max}}]\times[z_{\mathrm{r}}^{\mathrm{min}},z_{\mathrm{r}}^{\mathrm{max}}]$. 

\subsection{Channel Model}
In this paper, quasi-static channels are assumed for both the source-to-relay (S-R) and relay-to-destination (R-D) links, where the channel vectors vary only with the MA positions. To model the channel in terms of the antenna positions, we adopt the field-response model~\cite{zhu2024MA}, where the channel response is modeled as the superposition of multiple paths. Specifically, denote the number of transmit and receive paths of the S-R link by $L^{\mathrm{sr,t}}_{k}$ and $L^{\mathrm{sr,r}}_{k}$, respectively, and denote the number of transmit and receive paths of the R-D link by $L^{\mathrm{rd,t}}_{k}$ and $L^{\mathrm{rd,r}}_{k}$. Then, for the S-R link, denote the elevation and azimuth AoAs of the $\ell$-th path as $\theta^{\mathrm{sr,r}}_{k,\ell}$ and $\phi^{\mathrm{sr,r}}_{k,\ell}, 1 \leq \ell \leq L^{\mathrm{sr,r}}_{k},1 \leq k \leq K $, respectively. Accordingly, the receive field-response vector (FRV) between the $k$-th source user and the $m$-th MA at the $n$-th time slot is given by
\begin{equation}
	\mathbf{f}_{k}(\mathbf{q}_m[n]) = \left[ e^{\mathrm{j} \frac{2\pi}{\lambda} \rho_{1,k}^{\mathrm{sr,r}}(\mathbf{q}_m[n])}, \dots, e^{\mathrm{j} \frac{2\pi}{\lambda} \rho_{L^{\mathrm{sr,r}}_{k},k}^{\mathrm{sr,r}}(\mathbf{q}_m[n])} \right]^\mathrm{T},
\end{equation}
where $\rho_{k,\ell}^{\mathrm{sr,r}}(\mathbf{q}_m[n])$ denotes the propagation distance difference between the $m$-th MA and the reference point $\mathbf{o}=[0,0,0]^{\mathrm{T}}$ in the S-R link, and $\lambda$ is the carrier wavelength. 

By defining the virtual AoAs of the S-R link as
\begin{subequations}
	\begin{align}
	\vartheta_{k,\ell}^{\mathrm{sr,r}} &= \cos \theta_{k,\ell}^{\mathrm{sr,r}} \sin \phi_{k,\ell}^{\mathrm{sr,r}}, \\
	\varphi_{k,\ell}^{\mathrm{sr,r}} &= \sin \theta_{k,\ell}^{\mathrm{sr,r}}, \\
	\omega_{k,\ell}^{\mathrm{sr,r}} &= \cos \theta_{k,\ell}^{\mathrm{sr,r}} \cos \phi_{k,\ell}^{\mathrm{sr,r}},
	\end{align}
\end{subequations} 
the propagation distance difference $\rho_{k,\ell}^{\mathrm{sr,r}}(\mathbf{q}_m[n])$ is given by
\begin{equation}
	\rho^{\mathrm{sr,r}}_{k,\ell}(\mathbf{q}_m[n]) = x_m[n]\vartheta_{k,\ell}^{\mathrm{sr,r}} + y_m[n]\varphi_{k,\ell}^{\mathrm{sr,r}} + z_m[n]\omega_{k,\ell}^{\mathrm{sr,r}}.
\end{equation}

Similarly, denote the elevation and azimuth AoDs of the $\ell$-th path in the S-R link as $\theta^{\mathrm{sr,t}}_{k,\ell}$ and $\phi^{\mathrm{sr,t}}_{k,\ell}, 1 \leq \ell \leq L^{\mathrm{sr,t}}_{k},1 \leq k \leq K $, respectively. The corresponding transmit FRV between the $k$-th source user and the $m$-th MA is given by
\begin{equation}
\mathbf{g}(\mathbf{d}_k^{\mathrm{s}}) = \left[ e^{\mathrm{j} \frac{2\pi}{\lambda} \rho_{1}^{\mathrm{sr,t}}(\mathbf{d}_k^{\mathrm{s}})}, \dots, e^{\mathrm{j} \frac{2\pi}{\lambda} \rho_{L^{\mathrm{sr,t}}_{k}}^{\mathrm{sr,t}}(\mathbf{d}_k^{\mathrm{s}})} \right]^\mathrm{T},
\end{equation}
where $\mathbf{d}_k^{\mathrm{s}}=[x_{k}^{\mathrm{s}},y_{k}^{\mathrm{s}},z_{k}^{\mathrm{s}}]^{\mathrm{T}}$ denote the positions of the $k$-th source user. By defining the virtual AoDs of the S-R link as
\begin{subequations}
	\begin{align}
	\vartheta_{k,\ell}^{\mathrm{sr,t}} &= \cos \theta_{k,\ell}^{\mathrm{sr,t}} \sin \phi_{k,\ell}^{\mathrm{sr,t}}, \\
	\varphi_{k,\ell}^{\mathrm{sr,t}} &= \sin \theta_{k,\ell}^{\mathrm{sr,t}}, \\
	\omega_{k,\ell}^{\mathrm{sr,t}} &= \cos \theta_{k,\ell}^{\mathrm{sr,t}} \cos \phi_{k,\ell}^{\mathrm{sr,t}},
	\end{align}
\end{subequations} 
the propagation distance difference $\rho_{\ell}^{\mathrm{sr,t}}(\mathbf{d}_k^{\mathrm{s}})$ is given by
\begin{equation}
\rho_{\ell}^{\mathrm{sr,t}}(\mathbf{d}_k^{\mathrm{s}}) = -x_{k}^{\mathrm{s}}\vartheta_{k,\ell}^{\mathrm{sr,t}} - y_{k}^{\mathrm{s}}\varphi_{k,\ell}^{\mathrm{sr,t}} + z_{k}^{\mathrm{s}}\omega_{k,\ell}^{\mathrm{sr,t}}.
\end{equation}

For the R-D link, denote the elevation and azimuth AoDs of the $\ell$-th transmit path as $\theta^{\mathrm{rd,t}}_{\ell}$ and $\phi^{\mathrm{rd,t}}_{\ell}, 1 \leq \ell \leq L^{\mathrm{rd,t}}_{k}$, respectively. Accordingly, the transmit field-response vector (FRV) between the $m$-th MA and the $k$-th destination user for the R-D link at the $n$-th time slot is given by
\begin{equation}
\mathbf{g}_{k}(\mathbf{q}_m[n]) = \left[ e^{\mathrm{j} \frac{2\pi}{\lambda} \rho_{1}^{\mathrm{rd,t}}(\mathbf{q}_m[n])}, \dots, e^{\mathrm{j} \frac{2\pi}{\lambda} \rho_{L^{\mathrm{rd,t}}_{k}}^{\mathrm{rd,t}}(\mathbf{q}_m[n])} \right]^\mathrm{T},
\end{equation}
where $\rho_{k,\ell}^{\mathrm{rd,t}}(\mathbf{q}_m[n])$ denotes the propagation distance difference in the R-D link. By defining the virtual AoDs of the R-D link as
\begin{subequations}
	\begin{align}
	\vartheta_{k,\ell}^{\mathrm{rd,t}} &= \cos \theta_{k,\ell}^{\mathrm{rd,t}} \sin \phi_{k,\ell}^{\mathrm{rd,t}}, \\
	\varphi_{k,\ell}^{\mathrm{rd,t}} &= \sin \theta_{k,\ell}^{\mathrm{rd,t}}, \\
	\omega_{k,\ell}^{\mathrm{rd,t}} &= \cos \theta_{k,\ell}^{\mathrm{rd,t}} \cos \phi_{k,\ell}^{\mathrm{rd,t}},
	\end{align}
\end{subequations} 
the propagation distance difference $\rho_{k,\ell}^{\mathrm{rd,t}}(\mathbf{q}_m[n])$ is given by
\begin{equation}
	\rho^{\mathrm{rd,t}}_{k,\ell}(\mathbf{q}_m[n]) = -x_m[n]\vartheta_{k,\ell}^{\mathrm{rd,t}} - y_m[n]\varphi_{\ell}^{\mathrm{rd,t}} + z_m[n]\omega_{k,\ell}^{\mathrm{rd,t}}.
\end{equation}

Denote the elevation and azimuth AoAs of the $\ell$-th path in the R-D link as $\theta^{\mathrm{rd,r}}_{k,\ell}$ and $\phi^{\mathrm{rd,r}}_{k,\ell}, 1 \leq \ell \leq L^{\mathrm{rd,r}}_{k},1 \leq k \leq K $, respectively, and denote the positions of the $k$-th destination user as $\mathbf{d}_k^{\mathrm{d}}=[x_{k}^{\mathrm{d}},y_{k}^{\mathrm{d}},z_{k}^{\mathrm{d}}]^{\mathrm{T}}$. The receive FRV between the $m$-th MA and the $k$-th destination user is given by
\begin{equation}
\mathbf{f}(\mathbf{d}_k^{\mathrm{d}}) = \left[ e^{\mathrm{j} \frac{2\pi}{\lambda} \rho_{1}^{\mathrm{rd,r}}(\mathbf{d}_k^{\mathrm{d}})}, \dots, e^{\mathrm{j} \frac{2\pi}{\lambda} \rho_{L^{\mathrm{rd,r}}_{k}}^{\mathrm{rd,r}}(\mathbf{d}_k^{\mathrm{d}})} \right]^\mathrm{T},
\end{equation}
By defining the virtual AoDs of the R-D link as
\begin{subequations}
	\begin{align}
	\vartheta_{k,\ell}^{\mathrm{rd,r}} &= \cos \theta_{k,\ell}^{\mathrm{rd,r}} \sin \phi_{k,\ell}^{\mathrm{rd,r}}, \\
	\varphi_{k,\ell}^{\mathrm{rd,r}} &= \sin \theta_{k,\ell}^{\mathrm{rd,r}}, \\
	\omega_{k,\ell}^{\mathrm{rd,r}} &= \cos \theta_{k,\ell}^{\mathrm{rd,r}} \cos \phi_{k,\ell}^{\mathrm{rd,r}},
	\end{align}
\end{subequations} 
the propagation distance difference $\rho_{\ell}^{\mathrm{rd,r}}(\mathbf{d}_k^{\mathrm{d}})$ is given by
\begin{equation}
\rho_{\ell}^{\mathrm{rd,r}}(\mathbf{d}_k^{\mathrm{d}}) = x_{k}^{\mathrm{d}}\vartheta_{k,\ell}^{\mathrm{rd,r}} + y_{k}^{\mathrm{d}}\varphi_{k,\ell}^{\mathrm{rd,r}} + z_{k}^{\mathrm{d}}\omega_{k,\ell}^{\mathrm{rd,r}}.
\end{equation}

As such, the channel vector between the $k$-th source user and the relay node in the $n$-th time slot is obtained as
\begin{equation}
\mathbf{h}^{\mathrm{sr}}_{k}[n] = \mathbf{F}^{\mathrm{H}}_{k}(\widetilde{\mathbf{q}}[n]) \boldsymbol{\Sigma}^{\mathrm{sr}}_{k} \mathbf{g}(\mathbf{d}_k^{\mathrm{s}}),
\end{equation}
where $\mathbf{F}_{k}(\widetilde{\mathbf{q}}[n]) = \left[\mathbf{f}_{k}(\mathbf{q}_1[n]),\mathbf{f}_{k}(\mathbf{q}_2[n]),\dots,\mathbf{f}_{k}(\mathbf{q}_M[n])\right] \in \mathbb{C}^{L^{\mathrm{sr,r}}_{k} \times M} $ is the receive field-response matrix (FRM), $\boldsymbol{\Sigma}^{\mathrm{sr}}_{k} \in \mathbb{C}^{L^{\mathrm{sr,r}}_{k} \times L^{\mathrm{sr,t}}_{k}}$ represents the path-response matrix (PRM) between the reference transmit position and receive position.

Similarly, the channel vector between the relay node and the $k$-th destination user in the $n$-th time slot is obtained as
\begin{equation}
\mathbf{h}^{\mathrm{rd}}_{k}[n] = \left[\mathbf{f}(\mathbf{d}_k^{\mathrm{d}})^\mathrm{H} \boldsymbol{\Sigma}^{\mathrm{rd}}_{k} \mathbf{G}_{k}(\widetilde{\mathbf{q}}[n])\right]^{\mathrm{T}},
\end{equation}
where $\mathbf{G}_{k}(\widetilde{\mathbf{q}}[n]) = \left[\mathbf{g}_k(\mathbf{q}_1[n]),\mathbf{g}_k(\mathbf{q}_2[n]),\dots,\mathbf{g}_k(\mathbf{q}_M[n])\right] \in \mathbb{C}^{L^{\mathrm{rd,t}}_{k} \times M} $ is the transmit field-response matrix (FRM), and $\boldsymbol{\Sigma}^{\mathrm{rd}}_{k} \in \mathbb{C}^{L^{\mathrm{rd,r}}_{k} \times L^{\mathrm{rd,t}}_{k}}$ represents the path-response matrix (PRM).

Hereby, the S-R and R-D channel matrices in the $n$-th time slot are given by
\begin{subequations}
	\begin{align}
	\mathbf{H}_{\mathrm{sr}}[n] 
	&= [\mathbf{h}_{1}^{\mathrm{sr}}[n], \mathbf{h}_{2}^{\mathrm{sr}}[n], \cdots, \mathbf{h}_{K}^{\mathrm{sr}}[n]], \\[1mm]
	\mathbf{H}_{\mathrm{rd}}[n] 
	& = [\mathbf{h}_{1}^{\mathrm{rd}}[n], \mathbf{h}_{2}^{\mathrm{rd}}[n], \cdots, \mathbf{h}_{K}^{\mathrm{rd}}[n]].	
	\end{align}
\end{subequations}

\subsection{Signal Model}
In the MAC stage, the transmit signals of the $K$ source users are denoted by $\mathbf{x}[n]=[x_{1}[n],x_{2}[n],\cdots,x_{K}[n]]^{\mathrm{T}}$, with $\mathbb{E}[\mathbf{x}[n]\mathbf{x}[n]^H]=\mathbf{I}_K$. Then, the received signals at the relay node can be given by
\begin{equation}
\mathbf{y}_{\mathrm{sr}}[n] = \mathbf{H}_{\mathrm{sr}}[n]\mathbf{P}_{\mathrm{s}}[n]\mathbf{x}[n] + \mathbf{n}_\mathrm{sr}[n],
\end{equation}
where $\mathbf{P}_{\mathrm{s}}[n] = \mathrm{diag}(\sqrt{p^{\mathrm{sr}}_1[n]},\sqrt{p^{\mathrm{sr}}_2[n]},\cdots,\sqrt{p^{\mathrm{sr}}_K[n]})$ is the diagonal transmit power allocation matrix of the source users, and $\mathbf{n}_\mathrm{sr}[n]\sim\mathcal{CN}(0,\delta^2\mathbf{I}_M)$ denotes the complex additive white Gaussian noise (AWGN).

The receive combining matrix in the $n$-th time slot is denoted by $\mathbf{W}[n] = [\mathbf{w}_1[n],\mathbf{w}_2[n],\cdots,\mathbf{w}_K[n]]$. The signals at the relay node after receive combining are given by $\hat{\mathbf{x}}[n]=\mathbf{W}[n]^{\mathrm{H}} \mathbf{y}_{\mathrm{sr}}[n]$. Therefore, the combining output for the $k$-th user is given by
\begin{equation}
	\begin{aligned}
	\widehat{x}_{k}[n] 
	& = \sqrt{p_{k}^{\mathrm{sr}}[n]} \mathbf{w}_{k}[n]^{\mathrm{H}} \mathbf{h}_{k}^{\mathrm{sr}}[n] x_{k}[n] \\
	& + \mathbf{w}_{k}[n]^{\mathrm{H}} \sum_{j=1, j \neq k}^{K}  \sqrt{p_{j}^{\mathrm{sr}}[n]} \mathbf{h}_{j}^{\mathrm{sr}}[n] x_{j}[n] \\
	& + \mathbf{w}_{k}[n]^{\mathrm{H}} \mathbf{n}_{\mathrm{sr}}[n]
	\end{aligned}.
\end{equation}

Since the relay node adopts the DF protocol, the transmit signal for the BC stage is regenerated from $\hat{\mathbf{x}}[n]$. For simplicity, perfect decoding is assumed at the relay, i.e., $\hat{\mathbf{x}}[n]$ can be correctly mapped to $\mathbf{x}[n]$~\cite{wi2011DF}. Then, the transmit signal is given by $\mathbf{s}[n]=\mathbf{V}[n]\mathbf{x}[n]$, where $\mathbf{V}[n]=[\mathbf{v}_1[n],\mathbf{v}_2[n],\ldots,\mathbf{v}_K[n]]$ denotes the transmit beamforming matrix in the $n$-th time slot.
Similarly, we can obtain the received signal at the $k$-th destination user
\begin{equation}
	\begin{aligned}
	y_{k}^{\mathrm{rd}}[n] & = \mathbf{h}_{k}^{\mathrm{rd}}[n]^{\mathrm{H}} \mathbf{v}_{k}[n] x_{k}[n]\\
	& + \mathbf{h}_{k}^{\mathrm{rd}}[n]^{\mathrm{H}} \sum_{j=1, j \neq k}^{K} \mathbf{v}_{j}[n] x_{j}[n] + n_{k}^{\mathrm{rd}}[n],
	\end{aligned}
\end{equation}
where $n_{k}^{\mathrm{rd}}[n] \sim \mathcal{CN}(0,\delta^2)$ represents the AWGN.

Then, the SINRs over the S–R and R–D links are given by
\begin{subequations}
	\begin{align}
	\gamma_{k}^{\mathrm{sr}}[n] &= \frac{ p_{k}^{\mathrm{sr}}[n] \left| \mathbf{w}_{k}[n]^{\mathrm{H}} \mathbf{h}_{k}^{\mathrm{sr}}[n] \right|^2 }{ \sum\limits_{j=1, j \neq k}^{K} p_{j}^{\mathrm{sr}}[n] \left| \mathbf{w}_{k}[n]^{\mathrm{H}} \mathbf{h}_{j}^{\mathrm{sr}}[n] \right|^2 + \delta^2\left\| \mathbf{w}_{k}[n] \right\|^2  },  \\
	\gamma_{k}^{\mathrm{rd}}[n] &= \frac{ \left| \mathbf{h}_{k}^{\mathrm{rd}}[n]^{\mathrm{H}} \mathbf{v}_{k}[n] \right|^2 }{ \sum\limits_{j=1, j \neq k}^{K} \left| \mathbf{h}_{k}^{\mathrm{rd}}[n]^{\mathrm{H}} \mathbf{v}_{j}[n] \right|^2 + \delta^2 }. 
	\end{align}
\end{subequations}
Therefore, the achievable rates of the $k$-th user pair over the S–R and R–D links are given by
\begin{subequations}
	\begin{align}
	R_{k}^{\mathrm{sr}}[n] &= \mathrm{log}_2 \left( 1+ \gamma_{k}^{\mathrm{sr}}[n]	\right),  \\
	R_{k}^{\mathrm{rd}}[n] &= \mathrm{log}_2 \left( 1+ \gamma_{k}^{\mathrm{rd}}[n]	\right). 
	\end{align}
\end{subequations}

\begin{figure}[t]
	\begin{center}
		\includegraphics[width=6.5 cm]{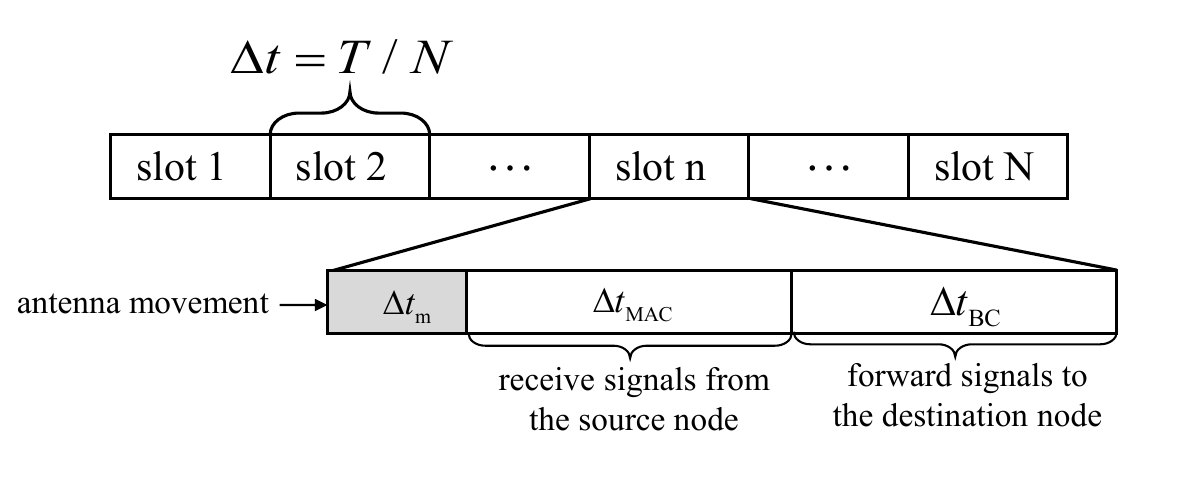}
		\caption{Three subslots for the antenna movement, the information reception and transmission at the relay.}
		\label{fig:slot}
	\end{center}
	\vspace{-0.25cm}
\end{figure}

\subsection{Relay Model}
Based on the achievable rates of the two links, the end-to-end relay transmission process is described as follows. As shown in Fig. \ref{fig:slot}, we consider a time slot structure consisting of three subslots to accommodate antenna movement and the two-stage relay transmission. Specifically, in the $n$-th time slot, the MAs first adjust their positions during the first subslot. The relay node then receives and decodes the signals transmitted by the source users and stores the decoded data in the buffer during the second subslot. During the third subslot, the buffered data is forwarded to the destination users.

Let $c_k[n]$ denote the buffer occupancy of the $k$-th user pair before the MAC stage in the $n$-th time slot. When the buffer capacity is sufficient, the source users will transmit at the maximum rate; otherwise, each source user reduces its transmission rate to avoid exceeding the remaining buffer capacity. Consequently, the effective transmission rate is given by
\begin{equation}
R_{k}^{\mathrm{MAC}}[n] = 
\begin{cases} 
R_{k}^{\mathrm{sr}}[n], \sum \limits_{k=1}^{K} (c_k[n]+R_{k}^{\mathrm{sr}}[n]\Delta t_{\mathrm{MAC}}) \leq C_{\mathrm{max}}, \\
\dfrac{C_{\mathrm{max}}-\sum \limits_{k=1}^{K}c_k[n]}{\sum \limits_{k=1}^{K}R_{k}^{\mathrm{sr}}[n]\Delta t_{\mathrm{MAC}}}R_{k}^{\mathrm{sr}}[n], \mathrm{otherwise}.
\end{cases}
\end{equation}
The buffer occupancy of the $k$-th user pair after the MAC stage is updated as
\begin{equation}
c_{k}^{\mathrm{MAC}}[n] = c_k[n] + R_{k}^{\mathrm{MAC}}[n]\Delta t_{\mathrm{MAC}}.
\end{equation}

Similarly, during the BC stage, the relay transmits at the achievable rate $R_k^{\mathrm{rd}}[n]$ only when the buffer contains sufficient data; otherwise, the relay node forwards all the stored information to the destination users. The effective transmission rate of the $k$-th user during the BC stage is given by
\begin{equation}
R_{k}^{\mathrm{BC}}[n] = 
\begin{cases} 
R_{k}^{\mathrm{rd}}[n], & R_{k}^{\mathrm{rd}}[n]\Delta t_{\mathrm{BC}} \leq c_{k}^{\mathrm{MAC}}[n], \\[2mm]
\dfrac{c_{k}^{\mathrm{MAC}}[n]}{\Delta t_{\mathrm{BC}}},& \mathrm{otherwise}.
\end{cases}
\end{equation}
Then, the buffer occupancy of the $k$-th user pair after BC at the $n$-th time slot is updated as
\begin{equation}
\begin{aligned}
c_{k}^{\mathrm{BC}}[n] &= c_{k}^{\mathrm{MAC}}[n] - R_{k}^{\mathrm{BC}}[n]\Delta t_{\mathrm{BC}}\\
                       &= c_{k}[n+1].
\end{aligned}
\end{equation}
In each subsequent time slot, the relay system repeats the above procedure, including antenna movement, the MAC stage, and the BC stage, until the final time slot is reached. Over the entire transmission period, the end-to-end transmission rate of the $k$-th user pair is determined by the effective transmission rate during the BC stage, i.e., $R_k[n] = R_{k}^{\mathrm{BC}}[n]$.

\subsection{Energy Model}
In this paper, the energy consumption for MA movement is taken into account. Each MA is assumed to be driven by stepper motors aligned with the X-axis, Y-axis, and Z-axis, respectively. The total movement distance of the $m$-th antenna along three axes in the $n$-th time slot is given by
\begin{equation}
\begin{aligned}
\Delta q_m[n] &= \Delta x_m[n] + \Delta y_m[n] + \Delta z_m[n]\\ &=\left|x_{m}[n]-x_{m}[n-1]\right|+\left|y_{m}[n]-y_{m}[n-1]\right|\\
&+\left|z_{m}[n]-z_{m}[n-1]\right|.
\end{aligned}
\end{equation}
We consider the flexible-position power model in~\cite{wu2025move} and~\cite{yang2026ee}, in which the MAs move at a constant speed $v_{\mathrm{MA}}$, and the movement energy consumption is modeled to be proportional to the movement distance. Thus, the movement energy consumption of the $m$-th MA in the $n$-th time slot is given by
\begin{equation}
E_{m}^{\mathrm{mov}}[n] = 
\begin{cases} 
0, & n = 1, \\
\eta\Delta q_m[n], & 2 \le n \le N,
\end{cases}
\end{equation}
where the constant $\eta$ denotes the energy consumption per unit movement distance. The total movement energy consumption of the relay node during the $n$-th time slot is given by
\begin{equation}
E_{\mathrm{mov}}[n] = \sum\limits_{m=1}^{M}E_{m}^{\mathrm{mov}}[n].
\end{equation}
In addition, the relay node allocates transmit power to the destination users during the BC stage. The corresponding transmission energy consumption during the $n$-th time slot is given by
\begin{equation}
E_{\mathrm{t}}[n] = \sum\limits_{k=1}^{K}\lVert \mathbf{v}_k[n]  \rVert_2^2\Delta t_{\mathrm{BC}}.
\end{equation}

\subsection{Problem Formulation}
To ensure fairness among $K$ user pairs, we aim to maximize the minimum average achievable rate of the considered multiuser two-link relay system. In each time slot, the antenna positions, transmit and receive beamforming matrices, and power allocation are jointly optimized while accounting for the energy tradeoff between antenna movement and signal transmission. Therefore, the optimization problem is formulated as
{
\fontsize{9.5pt}{11pt}\selectfont
\begin{subequations}\label{prob}
	\begin{align}
	&\mathop{\max}\limits_{\tilde{\mathbf{q}},\mathbf{P}_{\mathrm{s}},\mathbf{W},\mathbf{V}}
	 \min \limits_{k}\frac{1}{N}\sum \limits_{n=1}^{N}R_k[n]\label{c1}\\
	\mathrm{s.t.}~~~&\mathbf{q}_m[n] \in \mathcal{C}_{r}, 1 \leq m \leq M,1 \leq n \leq N,\label{c2}\\
	&\lVert \mathbf{q}_i[n]-\mathbf{q}_j[n]\rVert_2 \geq \frac{\lambda}{2}, 1 \leq n \leq N, 1\leq i < j \leq M,\label{c3}\\
	&\left|\Delta x_m[n]\right| \leq v_{\mathrm{MA}}\Delta t_{\mathrm{m}}, 2 \leq n \leq N, 1 \leq m \leq M,\label{c4}\\
	&\left|\Delta y_m[n]\right| \leq v_{\mathrm{MA}}\Delta t_{\mathrm{m}},1 \le m \le M, 2 \leq n \leq N,\label{c5}\\
	&\left|\Delta z_m[n]\right| \leq v_{\mathrm{MA}}\Delta t_{\mathrm{m}},1 \le m \le M, 2 \leq n \leq N,\label{c6}\\
	&0 \leq p^{\mathrm{sr}}_k[n] \leq P_{\mathrm{max}}^{\mathrm{s}},1 \le k \le K, 1 \leq n \leq N,\label{c7}\\
	&\sum\limits_{k=1}^{K}\lVert \mathbf{v}_k[n] \rVert_2^2 \leq P_{\mathrm{max}}^{\mathrm{r}}, 1 \leq n \leq N,\label{c8}\\
	&\sum\limits_{n=1}^{N} (E_{\mathrm{t}}[n] + E_{\mathrm{mov}}[n])\leq E_{\mathrm{total}},\label{c9} 
	\end{align}
\end{subequations}
}
where \eqref{c2} constrains the MAs to move within the predefined cuboid region. Constraint \eqref{c3} guarantees that the minimum inter-antenna spacing is no smaller than the half-wavelength. \eqref{c4}-\eqref{c6} constrain the maximum displacement of each MA along the X-axis, Y-axis, and Z-axis, respectively. \eqref{c7} limits the maximum transmit power of each source user. \eqref{c8} constrains the total transmit power of the relay node in each time slot. \eqref{c9} constrains the total energy budget of the relay node for antenna movement and signal transmission over the entire $N$ time slots.

It is worth noting that the MA positions, power allocation, and beamforming matrices are highly coupled in problem \eqref{prob}, making the problem non-convex and hard to solve by conventional mathematical methods. Furthermore, the long-term energy constraint and the sequential decision-making nature of the problem make it difficult to be addressed using traditional bio-inspired optimization algorithms. Therefore, we propose a TD3-based two-layer algorithm to obtain a suboptimal solution of problem \eqref{prob}.
\begin{figure*}[t]
	\vspace{0.25cm}
	\begin{center}
		\includegraphics[width=14 cm]{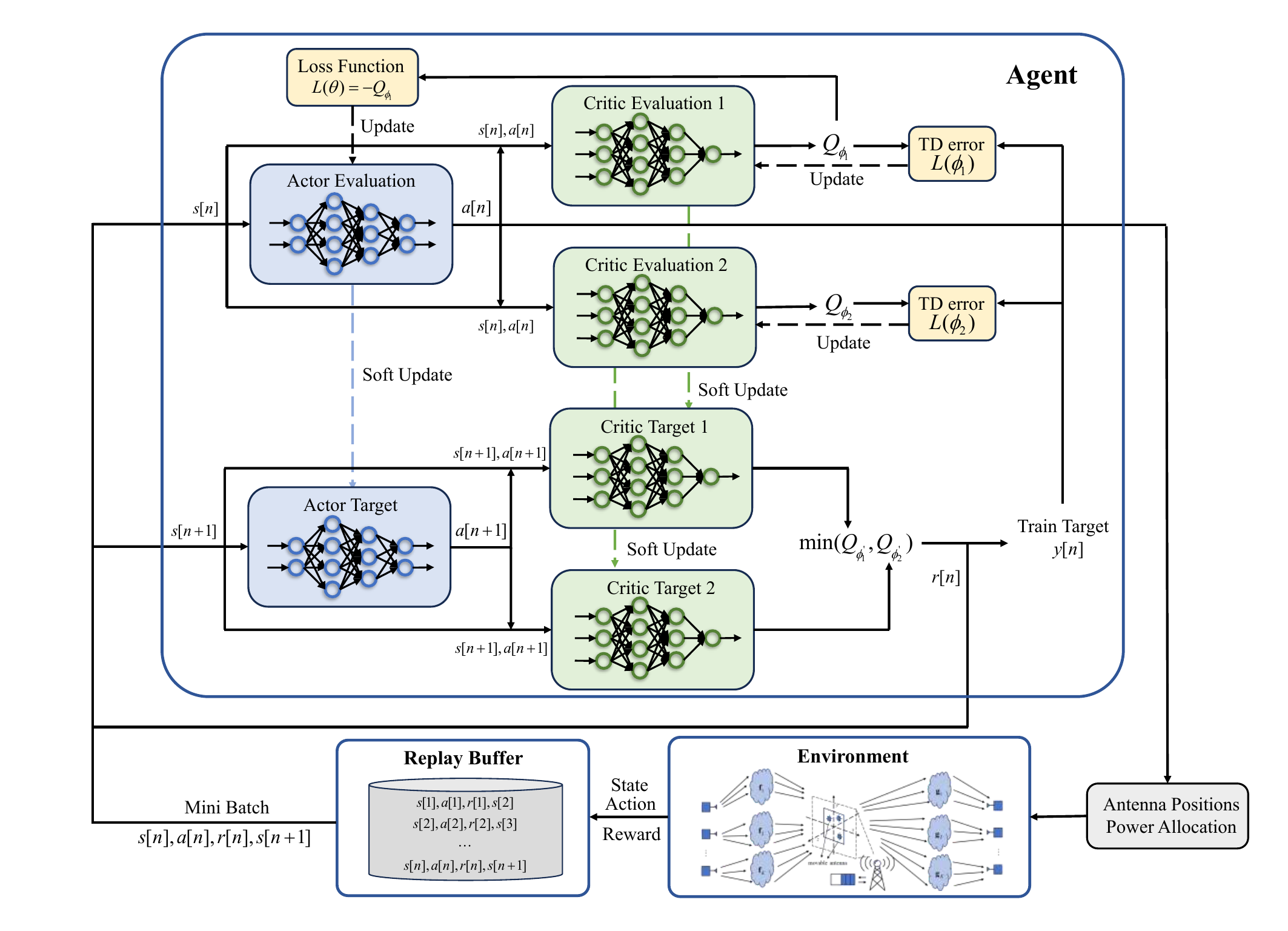}
		\caption{Illustration of the TD3 network structure.}
		\label{fig:TD3}
	\end{center}
\end{figure*}

\section{Proposed Solution}
To solve the formulated problem, we decompose the optimization problem into two layers in this section. In the outer layer, we employ a DRL algorithm to optimize the MA positions and the power allocation over all $N$ time slots. In the inner layer, the transmit and receive beamforming solution is obtained in closed form for each time slot. Specifically, in order to deal with the continuous optimization variables in the outer-layer problem, we adopt the TD3 algorithm, which alleviates overestimation bias and improves training stability compared with DDPG~\cite{fuji2018td3}.

\subsection{Partially Observable Markov Decision Process}
The DRL framework is formulated as a Markov decision process (MDP), which can be denoted by a quintuple $\left( S,A,P,R,\gamma \right)$. $S$ and $A$ in the MDP represent the state space and action space, respectively. $R$ denotes the reward function, which provides feedback to guide the agent during training. $P(s'|s,a)$ indicates the probability of the state transition from $s$ to $s'$ under the action $a$. $\gamma \in [0,1]$ denotes the discount factor.

In this paper, we utilize the TD3 algorithm to address the optimization of the antenna movement and the power allocation. The action space consists of the MA movement vector and uplink/downlink power allocation, i.e., $a[n] = \{ \{\Delta x_m[n],\Delta y_m[n],\Delta z_m[n]\},\{p^{\mathrm{sr}}_k[n]\},\{\tilde{p}^{\mathrm{rd}}_k[n]\},p_r[n]\}$, where $p_r[n]$ denotes the total transmit power of the relay node, and $\tilde{p}^{\mathrm{rd}}_k[n]$ represents the normalized power allocation ratio for the $k$-th destination user, i.e., $\tilde{p}^{\mathrm{rd}}_k[n] = \lVert \mathbf{v}_k[n] \rVert_2^2/p_r[n]$. To ensure the satisfaction of constraints \eqref{c4}-\eqref{c8}, the action space is designed by setting its upper and lower bounds to the feasible ranges of the corresponding variables in the constraints. The state space is designed to capture the environment information and is defined as $s[n] = \{\{\mathbf{q}_m[n]\},\{R_k[n-1]\},\{c_k[n]\},E_{\mathrm{res}}[n]\}$, where $E_{\mathrm{res}}[n]$ denotes the available energy in the $n$-th time slot and is given by
\begin{equation}
E_{\mathrm{res}}[n] = E_{\mathrm{total}}-\sum\limits_{i=1}^{n-1} \left(E_{\mathrm{t}}[i] + E_{\mathrm{mov}}[i]\right).
\end{equation}

Since constraints \eqref{c3} and \eqref{c9} cannot be directly enforced through the action space, we incorporate the corresponding penalty terms into the reward function. Constraint \eqref{c2} can be satisfied by applying action clipping to the movement variables. However, existing studies on DRL have shown that simple action clipping may lead to sample inefficiency and unstable training in tasks with high-dimensional action spaces~\cite{han2019drl}. Therefore, we also introduce an appropriate penalty term to discourage the agent from frequently generating invalid actions. Therefore, the reward function is designed as
\begin{equation} \label{reward}
r[n] = 
\begin{cases} 
-\alpha D[n], \eqref{c3}\ \mathrm{is\ not\ satisfied}, \\
\min \limits_{k}\frac{1}{n}\sum \limits_{i=1}^{n}R_k[i] - \beta_1 F_d[n] - \beta_2 F_p[n], \mathrm{otherwise},
\end{cases}
\end{equation}
where $\alpha$, $\beta_1,\beta_2>0$ are weighting factors. $\min \limits_{k}\sum_{i=1}^n\frac{1}{n}R_k[i]$ represents the cumulative objective value up to the $n$-th time slot. $D[n]$ denotes the number of MA pairs violating the minimum inter-antenna spacing constraint. $F_d[n]$ and $F_p[n]$ denote the penalty terms associated with violations of constraints \eqref{c2} and \eqref{c9}, respectively.

Let $\Delta \hat{x}_m[n]$, $\Delta \hat{y}_m[n]$, and $\Delta \hat{z}_m[n]$ denote the excess displacements of the unclipped movement actions beyond the prescribed region along the X-, Y-, and Z-axes, respectively. The penalty term $F_d[n]$ is defined as
\begin{equation}
F_d[n] = \sum\limits_{m=1}^{M}(\left|\Delta \hat{x}_m[n]\right|+\left|\Delta \hat{y}_m[n]\right|+\left|\Delta \hat{z}_m[n]\right|).
\end{equation}

Since constraint \eqref{c9} restricts the total energy consumption over the entire transmission period, we construct a time-slotted energy constraint based on the average energy budget as
\begin{equation}
\sum\limits_{i=1}^{n} (E_{\mathrm{t}}[i] + E_{\mathrm{mov}}[i])\leq n\frac{E_{total}}{N},1 \leq n \leq N.
\end{equation}
Therefore, the time-slotted penalty term $F_p[n]$ is defined as
\begin{equation}\label{punish1}
F_p[n] = \sum\limits_{i=1}^{n} (E_{\mathrm{t}}[i] + E_{\mathrm{mov}}[i]) - n\frac{E_{total}}{N},
\end{equation}
which allows the MAs to temporarily exceed the average energy budget when additional antenna movements can effectively improve the end-to-end transmission performance by enhancing the channel gain or reducing channel correlation, while encouraging the agent to save energy when further movements yield only limited rate improvement.

\subsection{TD3-based Algorithm}
As shown in Fig.~\ref{fig:TD3}, the TD3 framework consists of an agent, an environment, and an experience replay buffer. The agent in TD3 adopts an actor-critic neural network architecture. Specifically, the agent consists of an actor network and two critic networks, each of which has a corresponding target network. The online networks are updated through interactions with the environment, while the target networks provide stable targets for network updates. The actor network takes the state $s[n]$ as input and outputs the action $a[n]=\mu_{\theta}(s[n])$, where $\theta$ denotes the actor network parameters. The critic networks approximate the Q-functions to evaluate the state-action pairs. Specifically, the parameters of the two critic networks are denoted by $\phi_1$ and $\phi_2$, and their outputs are given by $Q_{\phi_1}(s[n],\mu_{\theta}(s[n]))$ and $Q_{\phi_2}(s[n],\mu_{\theta}(s[n]))$, respectively. The target actor network generates the next action $a[n+1]$ according to the next state $s[n+1]$, and the target critic networks evaluate the corresponding next state-action pair to obtain the target Q-values. The environment provides the next state and reward after executing each action. The state, action, and reward are stored in the replay buffer for updating the neural networks.

In the $n$-th time slot, the agent observes the current state $s[n]$ and generates an action $\tilde{a}[n]$ according to the actor evaluation network. To encourage exploration, Gaussian noise is added to the selected action before it is executed by the environment. Therefore, the actual action applied to the environment is expressed as
\begin{equation} \label{actionNoise}
a[n] = \mathrm{clip}(\tilde{a}[n] + \epsilon, a_{\mathrm{low}}, a_{\mathrm{high}}),
\end{equation}
where $\epsilon \sim \mathcal{N}(0,\sigma^2)$ is the Gaussian noise, and $\operatorname{clip}(\cdot)$ is the clipping function. $a_{\mathrm{low}}$ and $a_{\mathrm{high}}$ denote the lower and upper bounds of the action space, respectively, which determine the feasible ranges of the antenna movements and power allocation variables.
Next, the environment receives the action $a[n]$, transitions to the next state $s[n+1]$, and feeds back a reward function $r[n]$ to the agent. 
The reward is combined with the minimum target Q-value to construct the training target, where the discount factor $\gamma$ controls the contribution of future rewards in the target value estimation.
\begin{equation}
y[n] = r[n] + \gamma \min\limits_{i=1,2}Q_{\phi_{i}'}(s[n+1],a[n+1]).
\end{equation}
Then, the loss functions of critic evaluation networks are constructed by minimizing the squared temporal difference (TD) errors
\begin{subequations} \label{lossc}
	\begin{align}
	L(\phi_1) &= \mathbb{E}_{(s,a,r,s')}\left[\left(Q_{\phi_1}(s[n],a[n])-y[n]\right)^2\right],\\
	L(\phi_2) &= \mathbb{E}_{(s,a,r,s')}\left[\left(Q_{\phi_2}(s[n],a[n])-y[n]\right)^2\right].
	\end{align}
\end{subequations}
\begin{algorithm}[t]
	\label{TD3}
	\caption{The TD3 Algorithm for Antenna Positions and Power Allocation Optimization}
	\begin{algorithmic}[1]
		\STATE Initialize the parameters of the evaluation networks $\phi_i$, $\theta$, and the replay buffer
		\STATE Set the parameters of target networks $\phi_i' \leftarrow \phi_i$, $\theta' \leftarrow \theta$
		\STATE Initialize total training episode $N_{\mathrm{episode}}$, slots per episode $N$, and delay frequency $N_\mathrm{delay}$
		\FOR{$episode=1$ to $N_{\mathrm{episode}}$}
		\STATE Initialize the MA relay-enhanced communication environment
		\FOR{$step=1$ to $N$}
		\STATE Relay node observes its state $s$ from the environment and generates an action $a'$
		\STATE Add the action noise according to \eqref{actionNoise}
		\STATE Calculate the reward function according to \eqref{reward}
		\STATE Relay node receives the reward function $r$ and obtains the next state $s'$
		\STATE Store the $\{s,a,r,s'\}$ into the replay buffer
		\STATE Randomly sample a minibatch of the size $N_{\mathrm{batch}}$ from the replay buffer
		\STATE Update the critic evaluation network by minimizing the loss function $L(\phi_1)$ and $L(\phi_2)$
		\IF{$N \% N_{\mathrm{delay}} = 0$}
		\STATE Update the actor evaluation network by minimizing the loss function $L(\theta)$
		\STATE Update the parameters of the target network according to \eqref{updateTarget}
		\ENDIF
		\ENDFOR
		\ENDFOR
	\end{algorithmic}
\end{algorithm}

The expectation is calculated over the transitions sampled from the replay buffer.

In the actor-critic architecture, the actor network updates its parameters by maximizing the expected Q-value of the actions generated by the policy. In TD3 implementations, the loss function of the actor evaluation network is typically constructed by maximizing the first Q-function. Therefore, the loss function of the actor network is given by
\begin{equation} \label{lossa}
L(\theta) = -\mathbb{E}\left[Q_{\phi_1}(s[n],\mu_{\theta}(s[n]))\right].
\end{equation}
The parameters of the target networks are updated according to the soft-update manner:
\begin{subequations} \label{updateTarget}
	\begin{align}
	\phi_1' &= \tau \phi_1' + (1-\tau)\phi_1, \\
	\phi_2' &= \tau \phi_2' + (1-\tau)\phi_2, \\
	\theta' &= \tau \theta' + (1-\tau)\theta,
	\end{align}
\end{subequations}
where $\phi_1'$, $\phi_2'$, and $\theta'$ denote the parameters of the two critic target networks and the actor target network, respectively. $\tau$ is the soft update weighting coefficient. Notably, TD3 adopts a delayed update strategy, i.e., the actor network is updated only once after the critic network finishes $N_{\mathrm{delay}}$ updates.

The detailed procedure of the proposed algorithm is summarized in Algorithm 1. In lines 1-3, the system parameters and neural network parameters are initialized. In line 5, the environment containing the channel information and MA positions is initialized. In lines 7-11, the agent interacts with the environment and stores the experience samples into the replay buffer for the whole episode. In lines 12-13, experience data is sampled from the replay buffer, and two critic networks are updated according to \eqref{lossc}. In line 15, the actor network is updated according to \eqref{lossa} following the delayed policy update strategy. In line 16, the parameters of the three target networks are updated from the corresponding online networks according to the soft-update manner.

\subsection{Closed-form Beamforming}
In line 9 of Algorithm 1, the end-to-end rate calculation involved in the reward function requires the antenna positions, power allocation, and transmit/receive beamforming. While the DRL-based algorithm is employed to optimize the antenna positions and the power allocation, the transmit/receive beamforming design has not been addressed. Since the beamforming matrices are highly coupled with the antenna positions, it is difficult to obtain the globally optimal beamforming matrices ${\mathbf{W}[n]}$ and ${\mathbf{V}[n]}$ for all time slots. Therefore, for any given antenna position and power allocation, the beamforming optimization problem in the $n$-th time slot is formulated as
\begin{subequations}\label{probWV}
	\begin{align}
	\mathop{\max}\limits_{\mathbf{W}[n],\mathbf{V}[n]}
	& \min \limits_{k}R_k[n],\label{c01}\\
	&\sum\limits_{k=1}^{K}\lVert \mathbf{v}_k[n] \rVert_2^2 \leq P^{\mathrm{r}}_{\mathrm{max}}\label{c02}.
	\end{align}
\end{subequations}
Due to the buffer-aided relay architecture, the information transmission over the S-R and R-D links can be optimized separately. Specifically, the S-R link determines the data arrival rate and the amount of information accumulated in the relay buffer, while the R-D link directly affects the data delivery rate to destination users. Therefore, the transmit and receive beamforming optimization can be decomposed into two independent subproblems, which are detailed as follows.

For the receive combining optimization, we aim to maximize the minimum achievable rate of the S-R link, which improves the data arrival rate at the relay buffer and provides sufficient information for subsequent forwarding. Therefore, the receive beamforming optimization problem can be formulated as
\begin{equation}\label{probW}
	\mathop{\max}\limits_{\mathbf{W}[n]}
	\min \limits_{k}R_k^{\mathrm{sr}}[n].
\end{equation}
Since the achievable rate of each user in the MAC stage is determined by the corresponding S-R link rate, problem \eqref{probW} can be solved by maximizing the achievable rates of the S-R links. Specifically, the MMSE receiver is adopted to obtain the receive beamforming vectors, and the receive beamforming vector for the $k$-th user is given by
\begin{equation}
\mathbf{w}_k[n] = \left( \sum\limits_{j=1}^{K}p^{\mathrm{sr}}_j[n]\mathbf{h}_j^{\mathrm{sr}}[n]\mathbf{h}_j^{\mathrm{sr}}[n]^{\mathrm{H}}+\delta^2\mathbf{I}_M\right)^{-1}\mathbf{h}_k^{\mathrm{sr}}[n].
\end{equation}

For the R-D link, the achievable rate determines the data delivery capability of the relay and directly affects the end-to-end transmission performance. Therefore, the transmit beamforming optimization problem can be formulated as
\begin{subequations}\label{probV}
	\begin{align}
	\mathop{\max}\limits_{\mathbf{V}[n]}
	& \min \limits_{k}R_k^{\mathrm{rd}}[n],\\
	& \sum\limits_{k=1}^{K}\lVert \mathbf{v}_k[n] \rVert_2^2 \leq P^{\mathrm{r}}_{\mathrm{max}},
	\end{align}
\end{subequations}
which is difficult to solve since the transmit beamforming vectors of different users are coupled in the SINR expressions. According to~\cite{math}, the optimal beamforming structure can be given by
\begin{equation}
\hat{\mathbf{v}}_k[n] = \dfrac{\left(\mathbf{I}_{M}+\sum\limits_{j=1}^{K}\frac{\lambda_j}{\delta^2}\mathbf{h}_j^{\mathrm{rd}}[n](\mathbf{h}_j^{\mathrm{rd}}[n])^{\mathrm{H}}\right)^{-1}\mathbf{h}_k^{\mathrm{rd}}[n]}{\left\lVert \left(\mathbf{I}_{M}+\sum\limits_{j=1}^{K}\frac{\lambda_j}{\delta^2}\mathbf{h}_j^{\mathrm{rd}}[n](\mathbf{h}_j^{\mathrm{rd}}[n])^{\mathrm{H}}\right)^{-1}\mathbf{h}_k^{\mathrm{rd}}[n] \right\rVert_2},
\end{equation}
where Lagrange parameters $\lambda_j$ satisfy the sum property $\sum_{j=1}^{K}\lambda_j=p_r[n]$. Since directly optimizing $\lambda_j$ introduces additional complexity, we adopt a heuristic solution by setting all coefficients to the average value, i.e., $\lambda_j = p_r[n]/K$. The transmit power is provided by the TD3-based algorithm, and thus the suboptimal transmit beamforming vector for the $k$-th user can be obtained as
\begin{equation}
\mathbf{v}_k[n] = \sqrt{p^{\mathrm{rd}}_k}[n]\hat{\mathbf{v}}_k[n].
\end{equation}

\subsection{Computational Complexity Analysis}
We separate the computation costs into environment interaction and TD3 learning. Denote the number of floating-point operations (FLOPs) for each propagation of the actor network and critic network by $F_{\mathrm{a}}$ and $F_{\mathrm{c}}$, respectively. At each time slot, action generation requires one actor forward propagation, while channel calculation and closed-form beamforming incur $\mathcal{O}\{KM(L_{\mathrm{sr}}+L_{\mathrm{rd}})\}$ and $\mathcal{O}\{4KM^2+2M^3\}$ operations, respectively. Thus, the complexity of one environment interaction is
\begin{equation}
C_{\mathrm{env}}=
\mathcal{O} \{F_{\mathrm{a}}+KM(L_{\mathrm{sr}}+L_{\mathrm{rd}})+4KM^2+2M^3\}.
\end{equation}

For each time slot, TD3 updates the two critic networks, while the actor is updated every $N_{\mathrm{delay}}$ slots. Since backward propagation has the same order of complexity as forward propagation, the target networks and two critic updates require $\mathcal{O}\{N_{\mathrm{batch}}(F_{\mathrm{a}}+6F_{\mathrm{c}})\}$ operations per slot. The delayed actor update incurs an additional $\mathcal{O}\{N_{\mathrm{batch}}(2F_{\mathrm{a}}+2F_{\mathrm{c}})\}$ operations.Therefore, the per-slot learning complexity is
\begin{equation}
\begin{aligned}
C_{\mathrm{learn}}=&\mathcal{O}\{N_{\mathrm{batch}}(F_{\mathrm{a}}+6F_{\mathrm{c}})\}\\
&+\frac{1}{N_{\mathrm{delay}}}\mathcal{O}\{N_{\mathrm{batch}}(2F_{\mathrm{a}}+2F_{\mathrm{c}})\}.
\end{aligned}
\end{equation}

The training process consists of $N_{\mathrm{episode}}$ episodes, with $N$ time slots per episode. Therefore, the total computational complexity is 
\begin{equation}
C_{\mathrm{total}}=N_{\mathrm{episode}}N(C_{\mathrm{env}}+C_{\mathrm{learn}}).
\end{equation}

\begin{table}[t]
	\renewcommand{\arraystretch}{1.2}
	\caption{Simulation Parameters}
	\footnotesize
	\begin{center}
		\begin{tabular}{c|l|c}
			\hline
			\textbf{Parameter}
			&\multicolumn{1}{c|}{\textbf{Description}}                                 &\textbf{Value} \\
			\hline
			$P_{\mathrm{max}}^{\mathrm{s}}$ & \makecell[l]{Max transmission power for\\ each source user (dBm)} & $3$ \\ \hline
			$\lambda$ & Carrier wavelength (m) & $0.01$ \\ \hline
			$\eta$ & Movement power factor (J/m)  & $20$ \\ \hline
			$\Delta t_{\mathrm{m}}$ & Time duration for movement (ms) & $100$ \\ \hline
			$\Delta t_{\mathrm{MAC}}$ & Time duration for MAC (ms)  & $450$ \\ \hline
			$\Delta t_{\mathrm{BC}}$ & Time duration for BC (ms)  & $450$ \\ \hline
			$v_{\mathrm{MA}}$ & Movement velocity (m/s)  & $0.01$ \\ \hline
			$N_{\mathrm{episode}}$ & Number of training episodes  & $300$ \\ \hline
			$N$ & Slots per episode & $1000$ \\ \hline
			$N_{\mathrm{buffer}}$ & Replay buffer size & $4096$ \\ \hline
			$N_{\mathrm{batch}}$ & Random batch size & $128$ \\ \hline
			$N_{\mathrm{delay}}$ & Actor network update frequency & $4$ \\ \hline
			$\alpha$ & Punishment factor & $2$ \\ \hline
			$\beta_1$ & Punishment factor & $0.45$ \\ \hline
			$\beta_2$ & Punishment factor & $0.2$ \\ \hline
			$\gamma$ & Discount factor & $0.99$ \\ \hline
			$\alpha^{\mathrm{start}}_{\mathrm{a}}$ & Initial actor learning rate & $10^{-4}$ \\ \hline
			$\alpha^{\mathrm{start}}_{\mathrm{c}}$ & Initial critic learning rate & $10^{-3}$ \\ \hline
			$\tau$ & Soft update weighting & $0.95$ \\ \hline
		\end{tabular}
	\end{center}
\end{table}

\section{Simulation Results}
In this section, we conduct simulations to evaluate the effectiveness of the proposed algorithm for the MA relay-enhanced multiuser system. First, the simulation setup and benchmark schemes are illustrated, followed by the numerical results and discussions.

\subsection{Simulation Setup and Benchmark Schemes}
In this paper, without loss of generality, we consider the geometry channel model, where each transmit path is associated with a distinct receive path component. For simplicity, we assume identical numbers of transmit and receive paths between all users and the relay MAs, i.e., $L^{\mathrm{sr,t}}_{k}=L^{\mathrm{sr,r}}_{k}=L^{\mathrm{sr}}$, and $L^{\mathrm{rd,t}}_{k}=L^{\mathrm{rd,r}}_{k}=L^{\mathrm{rd}}$. The default number of paths is set to be $L^{\mathrm{sr}}=L^{\mathrm{rd}}=3$. The PRMs for two links are diagonal matrix, i.e., $\mathbf{\Sigma}_{\mathrm{sr}}=\mathrm{diag}\{\sigma^{\mathrm{sr}}_{1},\cdots,\sigma^{\mathrm{sr}}_{L_{\mathrm{sr}}}\}$ and $\mathbf{\Sigma}_{\mathrm{rd}}=\mathrm{diag}\{\sigma^{\mathrm{rd}}_{1},\cdots,\sigma^{\mathrm{rd}}_{L_{\mathrm{rd}}}\}$, where the elements $\sigma^\mathrm{sr}_{\ell} \sim \mathcal{CN}(0,1/L_{\mathrm{sr}})$ and $\sigma^\mathrm{rd}_{\ell} \sim \mathcal{CN}(0,1/L_{\mathrm{rd}})$ are CSCG variables, respectively. The elevation and azimuth AoAs and AoDs are randomly and uniformly distributed over $[-\pi/2,\pi/2]$. The moving region of the MAs is set as a cubic region $[-A/2,A/2]\times[-A/2,A/2]\times[-A/2,A/2]$, where $A=6\lambda$ by default. The number of communication user pairs is set to $K=4$, and both source users and destination users are randomly distributed on two planes parallel to the XOY plane. The number of MAs is set to $M=6$, which are initially deployed in the XOY plane with half-wavelength inter-antenna spacing. The total energy budget of the relay node is set to $E_{\mathrm{total}}=6$ J, and the maximal transmission power of the relay array is set to $P_{\mathrm{max}}^{\mathrm{r}} = 9$ dBm. The total buffer capacity is set to $C_{\mathrm{max}}=400$ bits. The actor network consists of $4$ fully connected layers, while the critic network consists of $3$ fully connected layers with $256$ neurons in each hidden layer. During the training stage, a learning-rate decay strategy is adopted for both the actor and the critic network, where the learning rate is multiplied by 0.9 for each 50 training episodes. The remaining system parameters and RL-related parameters are listed in Table I.

For performance comparison, we define the following benchmark schemes.
\begin{itemize}
	\item \textbf{Sparse-UPA}: The relay node is equipped with an FPA-based $2\times3$ array. The antennas are uniformly placed in the XOY plane with the maximum available aperture.
	\item \textbf{Dense-UPA}: The relay node is equipped with an FPA-based $2\times3$ array. The antennas are placed in the XOY plane with half-wavelength inter-antenna spacing.
	\item \textbf{Fixed MA}: The relay node is equipped with $M$ MAs, while the MA elements are fixed at the positions optimized by the PSO algorithm. In this case, the energy for antenna movement can be saved, while the continuous spatial DoFs cannot be fully exploited to enhance the relay communication performance.
	\item \textbf{MA w/o buffer}: The relay node is equipped with $M$ MAs and the antenna positions are optimized by the proposed TD3-based agorithm. However, the relay is not equipped with a buffer to temporarily store information received from the source users over multiple time slots, and the end-to-end transmission rate is always limited by the bottleneck link.	
\end{itemize}

\subsection{Numerical Results}
\begin{figure}[t]
	\vspace{0.25cm}
	\begin{center}
		\includegraphics[width=7.0 cm]{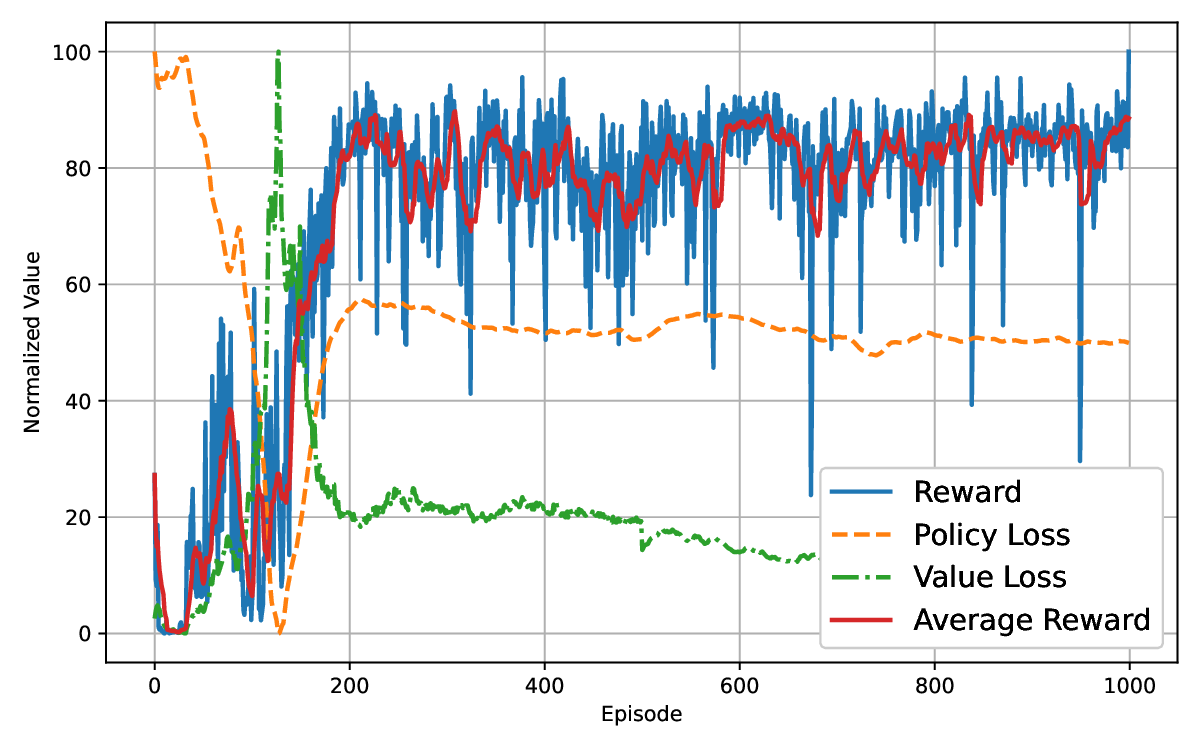}
		\caption{Convergence of rewards, actor loss, and critic loss functions.}
		\label{fig:train}
	\end{center}
	\vspace{-0.25cm}
\end{figure}

First, we examine the convergence performance of the proposed TD3 algorithm. The normalized reward, policy loss of the actor network, and Q-value loss of the critic networks are illustrated in Fig.~\ref{fig:train}. As can be observed, the reward increases with fluctuations during the first 200 episodes and then gradually converges to a stable value. The Q-value loss initially increases because the critic networks cannot accurately estimate the target Q-values in the early training stage, resulting in a larger TD error. As training progresses, the critic networks gradually learn a more accurate Q-function, and the loss decreases and finally stabilizes. In addition, the policy loss decreases from a relatively large value and eventually converges, which is consistent with the typical training behavior of the TD3 algorithm. These results demonstrate the effectiveness of the proposed TD3 algorithm for the MA relay-enhanced multiuser communication system.

\begin{figure}[t]
	\begin{center}
		\includegraphics[width=7.9 cm]{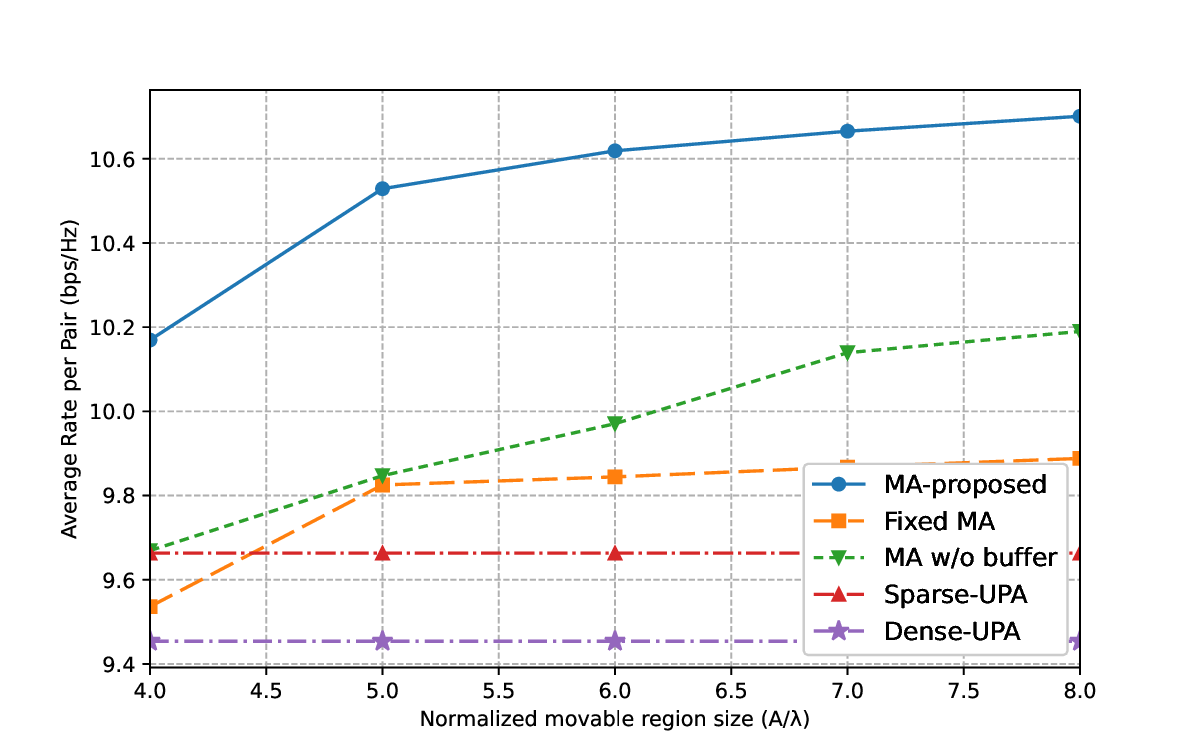}
		\caption{Average achievable rates for different schemes versus the movement region size.}
		\label{fig:size}
	\end{center}
	\vspace{-0.25cm}
\end{figure}

In Fig.~\ref{fig:size}, we compare the average achievable rates for different schemes versus the normalized movable region size. As can be observed, our proposed MA-based scheme achieves superior rate performance over all the benchmark schemes by benefiting from the flexible antenna position optimization and the buffer-aided mechanism at the relay node. On the one hand, by dynamically tuning antenna positions and leveraging the relay’s buffer capacity, both the source-relay and relay-destination channels can be adaptively configured to balance the transmission rates of the two-way links. On the other hand, flexible antenna repositioning substantially reduces channel correlation across different user pairs to achieve enhanced spatial multiplexing performance, thus effectively mitigating the interference between multiple user pairs and boosting the overall communication performance. In addition, the achievable rate of the proposed MA scheme is shown to increase monotonically with the normalized movable region size. This stems from the fact that a larger movable region offers antennas more spatial DoFs to fully exploit the continuous channel variations at the relay node.

\begin{figure}[t]
	\begin{center}
		\vspace{-0.1cm}
		\includegraphics[width=7.5 cm]{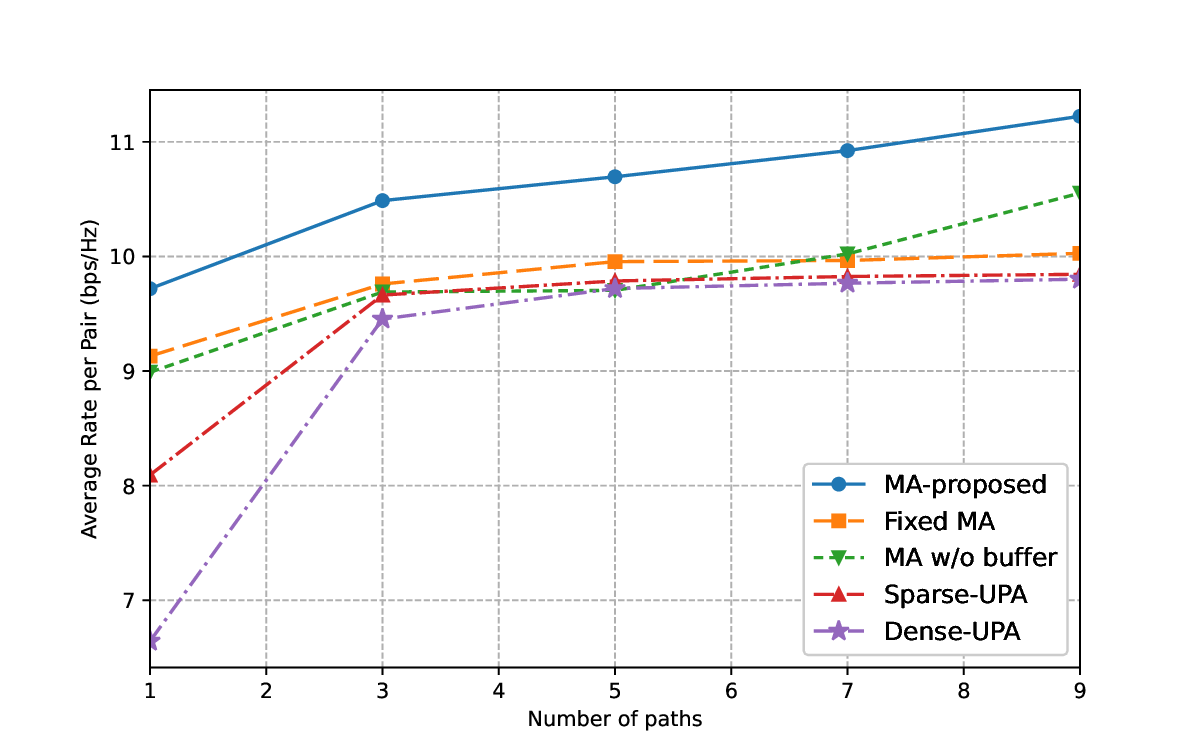}
		\caption{Average achievable rates for different schemes versus the number of
			paths.}
		\label{fig:path}
	\end{center}
	\vspace{-0.25cm}
\end{figure}

In Fig.~\ref{fig:path}, we compare the average achievable rates of different schemes versus the number of paths. It can still be observed that our proposed MA-based method outperforms all the benchmarks in terms of the achievable rate. It is shown that MA-based schemes achieve a higher achievable rate than both the sparse UPA and dense UPA methods by leveraging antenna movement to reduce channel correlation between different users. Furthermore, we can observe that the performance gap between the proposed MA-based solution and the ``Fixed MA" method increases with the number of paths, which indicates that the dynamic antenna movement of the proposed design enables more effective channel reconfiguration to achieve enhanced diversity and multiplexing performance in rich multi-path environments, thereby boosting the channel power gain and mitigating inter-pair interference.

\begin{figure}[t]
	\begin{center}
		\includegraphics[width=7.9 cm]{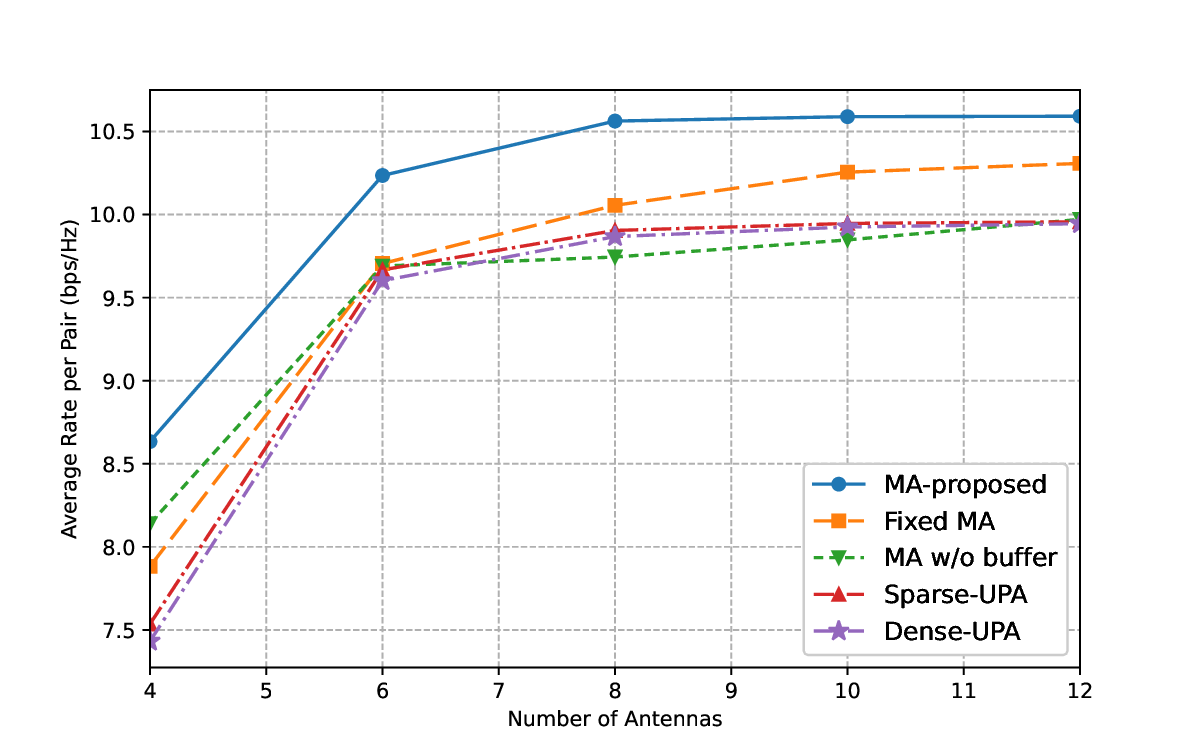}
		\caption{Average achievable rates for different schemes versus the number of
			movable antennas.}
		\label{fig:antenna}
	\end{center}
	\vspace{-0.35cm}
\end{figure}

In Fig.~\ref{fig:antenna}, we compare the average achievable rates of different schemes versus the number of antennas at the relay. As the number of antennas increases, all schemes achieve higher achievable rates, while the proposed scheme consistently outperforms other benchmark schemes. However, as the number of antennas becomes sufficiently large (e.g., $M=10$), the performance gain of the proposed scheme becomes less pronounced. This is because a larger number of MAs incurs higher antenna mobility-related energy consumption, which demonstrates the inherent rate-energy tradeoff in the MA relay system. Meanwhile, when the number of antennas becomes sufficiently large, the performance of the ``MA w/o buffer" scheme deteriorates significantly due to the energy consumption caused by antenna mobility, underperforming not only the ``Fixed MA" method but even the sparse and dense UPA-based benchmarks.

\begin{figure}[t]
	\begin{center}
		\vspace{-0.25cm}
		\includegraphics[width=7.9 cm]{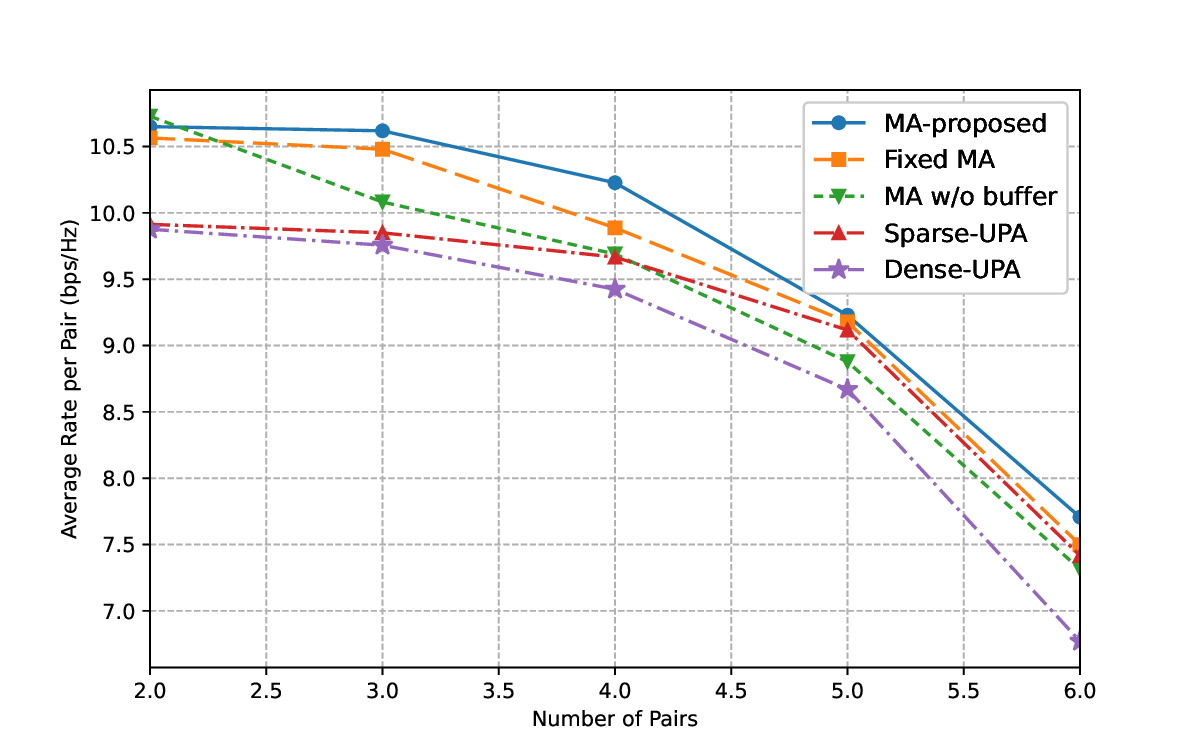}
		\caption{Average achievable rates for different schemes versus the number of
			pairs of users.}
		\label{fig:pair}
	\end{center}
	\vspace{-0.25cm}
\end{figure}

In Fig.~\ref{fig:pair}, we compare the average achievable rates of different schemes versus the number of user pairs. As the number of user pairs increases, the achievable rate of the proposed scheme decreases by $27.6\%$ compared with the case of only two user pairs, while our proposed solution always achieves the best performance compared to all the benchmark schemes. This performance degradation is mainly caused by the increasingly severe inter-user interference. In addition, a growing number of user pairs elevates the likelihood of rate imbalance between the S-R and R-D links. Consequently, it becomes more challenging for the buffer-aided mechanism to effectively coordinate data flows across the two-way links, resulting in a reduced performance gain from the buffer.

\begin{figure}[t]
	\begin{center}
		\includegraphics[width=7.5 cm]{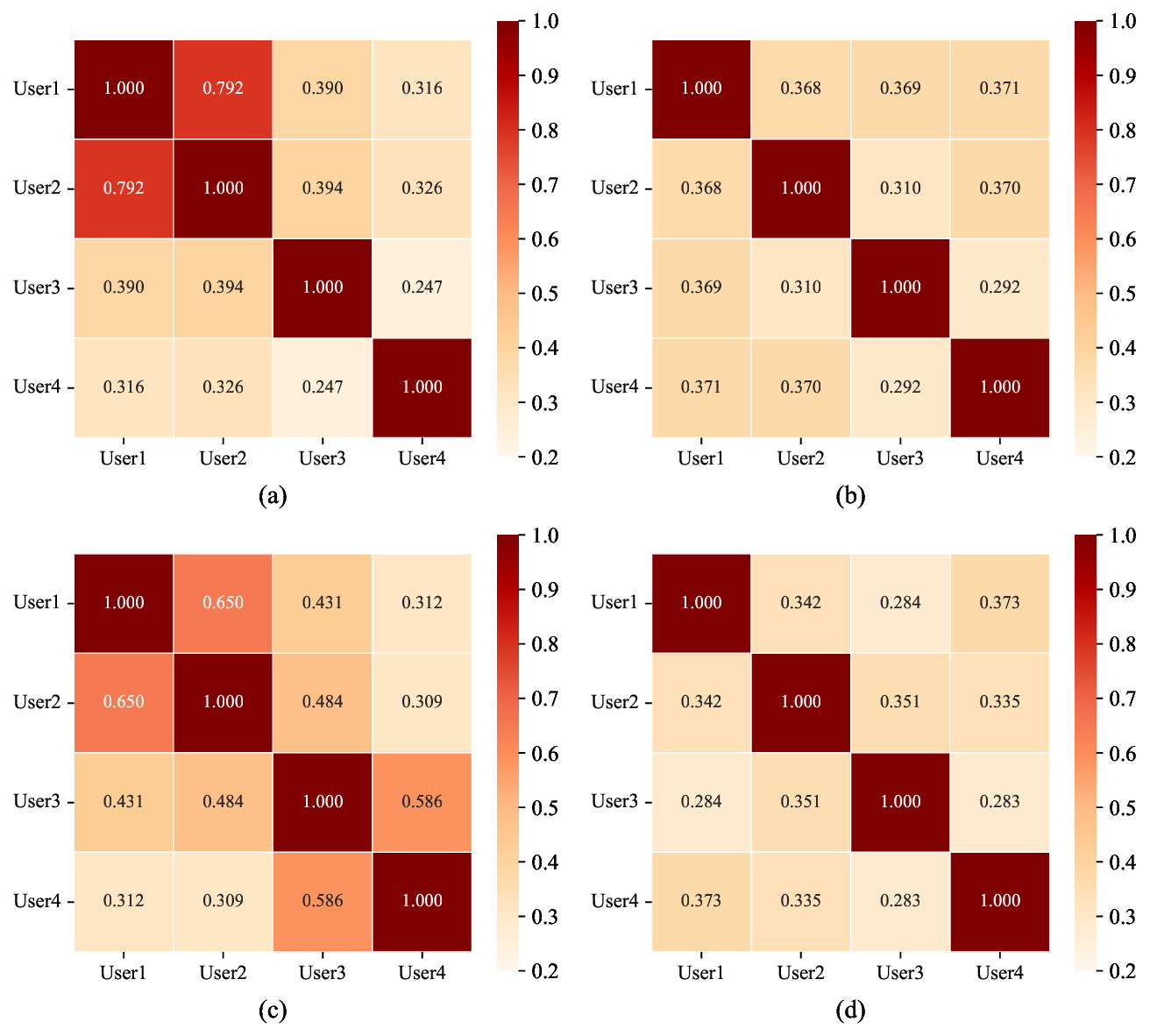}
		\caption{Inter-user channel correlation before and after optimization.}
		\label{fig:corr}
	\end{center}
	\vspace{-0.25cm}
\end{figure}

To obtain more insights, Fig.~\ref{fig:corr} illustrates the user channel correlation matrices of the S-R and R-D links before and after MA position optimization. Specifically, Figs.~\ref{fig:corr}(a) and \ref{fig:corr}(c) present the channel correlations of the S-R and R-D links when the MAs at the relay are fixed at their initial positions, whereas Figs.~\ref{fig:corr}(b) and \ref{fig:corr}(d) show the average channel correlations obtained over 1000 time slots with the proposed MA position optimization strategy. It can be observed that communication user pairs exhibit strong channel correlations in the initial antenna deployment, while the channel correlations are significantly reduced after MA position optimization. Moreover, it is observed that the channel correlation distribution becomes more balanced among different user pairs. This result demonstrates that the proposed algorithm can effectively mitigate undesired interferences among users, thereby achieving enhanced spatial multiplexing performance.

\section{Conclusion}
In this paper, we investigated an MA relay-enhanced multiuser communication system, where a half-duplex DF relay equipped with an MA array and a buffer was employed to enhance the end-to-end transmission performance. A general multipath field-response channel model was adopted to characterize the wireless channels corresponding to different MA positions. Based on this channel model, we formulated an optimization problem to maximize the minimum average achievable rate among all user pairs by jointly optimizing the MA positions, transmit power allocation, and transmit/receive beamforming matrices. To solve the non-convex optimization problem with highly coupled variables, we proposed an RL-based two-layer framework, where a TD3 algorithm was utilized to optimize the MA positions and power allocation, and the transmit and receive beamforming matrices were obtained through closed-form derivation in each time slot. Simulation results verified the effectiveness of the proposed algorithm and demonstrated the superiority of the proposed MA relay-enhanced multiuser communication system over conventional FPA-based relay systems and other MA-based benchmark schemes. Moreover, the buffer-aided mechanism effectively alleviated the rate imbalance between the S-R and R-D links, thereby enhancing the minimum average achievable rate among all users.

\bibliographystyle{IEEEtran} 
\bibliography{IEEEabrv,ref}

\begin{thebibliography}{10}
\providecommand{\url}[1]{#1}
\csname url@samestyle\endcsname
\providecommand{\newblock}{\relax}
\providecommand{\bibinfo}[2]{#2}
\providecommand{\BIBentrySTDinterwordspacing}{\spaceskip=0pt\relax}
\providecommand{\BIBentryALTinterwordstretchfactor}{4}
\providecommand{\BIBentryALTinterwordspacing}{\spaceskip=\fontdimen2\font plus
\BIBentryALTinterwordstretchfactor\fontdimen3\font minus
  \fontdimen4\font\relax}
\providecommand{\BIBforeignlanguage}[2]{{%
\expandafter\ifx\csname l@#1\endcsname\relax
\typeout{** WARNING: IEEEtran.bst: No hyphenation pattern has been}%
\typeout{** loaded for the language `#1'. Using the pattern for}%
\typeout{** the default language instead.}%
\else
\language=\csname l@#1\endcsname
\fi
#2}}
\providecommand{\BIBdecl}{\relax}
\BIBdecl

\bibitem{me2025MA}
Y.~Zhang, L.~Zhu, X.~Pi, H.~Mao, and Z.~Xiao, ``Reinforcement learning-based
  position optimization for movable antenna-enabled relay communications,'' in
  \emph{Proc. IEEE Int. Conf. Commun. Workshops}, Jun. 2025, pp. 56--61.

\bibitem{saad20206G}
W.~Saad, M.~Bennis, and M.~Chen, ``A vision of {6G} wireless systems:
  Applications, trends, technologies, and open research problems,'' \emph{IEEE
  Netw.}, vol.~34, no.~3, pp. 134--142, May 2020.

\bibitem{mimo2004}
A.~PAULRAJ, D.~GORE, R.~NABAR, and H.~BOLCSKEI, ``An overview of {MIMO}
  communications - a key to gigabit wireless,'' \emph{Proc. IEEE}, vol.~92,
  no.~2, pp. 198--218, Feb. 2004.

\bibitem{massive2014}
E.~G. Larsson, O.~Edfors, F.~Tufvesson, and T.~L. Marzetta, ``Massive {MIMO}
  for next generation wireless systems,'' \emph{IEEE Commun. Mag.}, vol.~52,
  no.~2, pp. 186--195, Feb. 2014.

\bibitem{massive2019}
M.~Wang, F.~Gao, S.~Jin, and H.~Lin, ``An overview of enhanced massive {MIMO}
  with array signal processing techniques,'' \emph{IEEE J. Sel. Top. Signal
  Process.}, vol.~13, no.~5, pp. 886--901, Sep. 2019.

\bibitem{rapp2019Thz}
T.~S. Rappaport, Y.~Xing, O.~Kanhere, S.~Ju, A.~Madanayake, S.~Mandal,
  A.~Alkhateeb, and G.~C. Trichopoulos, ``Wireless communications and
  applications above 100 ghz: Opportunities and challenges for {6G} and
  beyond,'' \emph{IEEE Access}, vol.~7, pp. 78\,729--78\,757, 2019.

\bibitem{rusek2013XLMIMO}
F.~Rusek, D.~Persson, B.~K. Lau, E.~G. Larsson, T.~L. Marzetta, O.~Edfors, and
  F.~Tufvesson, ``Scaling up {MIMO}: Opportunities and challenges with very
  large arrays,'' \emph{IEEE Signal Process. Mag.}, vol.~30, no.~1, pp. 40--60,
  Jan. 2013.

\bibitem{multiuser2004}
Q.~Spencer, C.~Peel, A.~Swindlehurst, and M.~Haardt, ``An introduction to the
  multi-user {MIMO} downlink,'' \emph{IEEE Commun. Mag.}, vol.~42, no.~10, pp.
  60--67, Oct. 2004.

\bibitem{mimo2009}
J.~Mietzner, R.~Schober, L.~Lampe, W.~H. Gerstacker, and P.~A. Hoeher,
  ``Multiple-antenna techniques for wireless communications - a comprehensive
  literature survey,'' \emph{IEEE Commun. Surveys Tuts.}, vol.~11, no.~2, pp.
  87--105, 2009.

\bibitem{Bj2016mimo}
E.~Björnson, E.~G. Larsson, and T.~L. Marzetta, ``Massive {MIMO}: ten myths
  and one critical question,'' \emph{IEEE Commun. Mag.}, vol.~54, no.~2, pp.
  114--123, Feb. 2016.

\bibitem{MA2024}
L.~Zhu, W.~Ma, and R.~Zhang, ``Movable antennas for wireless communication:
  Opportunities and challenges,'' \emph{IEEE Commun. Mag.}, vol.~62, no.~6, pp.
  114--120, Jun. 2024.

\bibitem{MA2026}
L.~Zhu, W.~Ma, W.~Mei, Y.~Zeng, Q.~Wu, B.~Ning, Z.~Xiao, X.~Shao, J.~Zhang, and
  R.~Zhang, ``A tutorial on movable antennas for wireless networks,''
  \emph{IEEE Commun. Surveys Tuts.}, vol.~28, pp. 3002--3054, 2026.

\bibitem{zhu2024FAS}
L.~Zhu and K.-K. Wong, ``Historical review of fluid antenna and movable
  antenna,'' \emph{arXiv preprint arXiv:2401.02362}, 2024.

\bibitem{zhu2026}
L.~Zhu, H.~Sun, W.~Ma, Z.~Xiao, and R.~Zhang, ``Multiuser communications aided
  by cross-linked movable antenna array: Architecture and optimization,''
  \emph{IEEE Trans. Wireless Commun.}, vol.~25, pp. 6729--6744, 2026.

\bibitem{xiu2026MA}
Y.~Xiu, Y.~Zhao, R.~Yang, W.~Lyu, D.~Niyato, D.~In~Kim, G.~Liu, and N.~Wei,
  ``Robust optimization for movable antenna-aided cell-free {ISAC} with time
  synchronization errors,'' \emph{IEEE Trans. Wireless Commun.}, vol.~25, pp.
  10\,082--10\,097, 2026.

\bibitem{amhaz2026CoMP}
A.~Amhaz, S.~Khisa, M.~Elhattab, C.~Assi, and S.~Sharafeddine, ``Enhancing
  {C}o{MP}-{RSMA} performance with movable antennas: A meta-learning
  optimization framework,'' \emph{IEEE Trans. Commun.}, vol.~74, pp.
  3802--3813, 2026.

\bibitem{liu2026NFC}
S.~Liu, X.~Yu, J.~Xu, and R.~Zhang, ``Near-field communication with massive
  movable antennas: A functional perspective,'' \emph{IEEE Trans. Wireless
  Commun.}, vol.~25, pp. 14\,455--14\,470, 2026.

\bibitem{chen2026MA}
X.~Chen, B.~Feng, Y.~Wu, D.~W.~K. Ng, and R.~Schober, ``Two-timescale sum-rate
  maximization for movable antenna enhanced systems,'' \emph{IEEE Trans.
  Wireless Commun.}, vol.~25, pp. 3894--3909, 2026.

\bibitem{zhu2026ToMA}
L.~Zhu, H.~Mao, W.~Ma, Z.~Xiao, J.~Zhang, and R.~Zhang, ``Towed movable antenna
  ({T}o{MA}) array for ultra secure airborne communications,'' \emph{IEEE J.
  Sel. Areas Commun.}, vol.~44, pp. 3766--3781, 2026.

\bibitem{ma2026ISAC}
W.~Ma, L.~Zhu, and R.~Zhang, ``Movable antenna enhanced integrated sensing and
  communication via antenna position optimization,'' \emph{IEEE Trans. Signal
  Process.}, vol.~74, pp. 1522--1537, 2026.

\bibitem{ke2026MA}
K.~Zhu, H.~Yin, T.~Wang, D.~Mishra, J.~Yuan, and Y.~Zhang, ``Movable
  antenna-based phased array: Beam pattern synthesis and experimental
  validations,'' \emph{IEEE Trans. Commun.}, vol.~74, pp. 10\,612--10\,626,
  2026.

\bibitem{yang2026MA}
S.~Yang, X.~Wei, N.~Su, W.~Mei, Z.~Chen, and B.~Ning, ``Movable
  antenna-enhanced near-field flexible beamforming: Performance analysis and
  optimization,'' \emph{IEEE Trans. Veh. Technol.}, vol.~75, no.~7, pp.
  13\,403--13\,418, Jul. 2026.

\bibitem{fu2025MA}
M.~Fu, L.~Zhu, and R.~Zhang, ``Extremely large-scale movable antenna-enabled
  multiuser communications: Modeling and optimization,'' \emph{IEEE Trans.
  Wireless Commun.}, vol.~25, pp. 10\,290--10\,304, 2026.

\bibitem{wang2025CFN}
Y.~Wang, C.~Han, S.~Sun, and J.~Zhang, ``Cross-far-and-near-field channel
  measurement and modeling in extremely large-scale movable antenna array
  systems,'' \emph{IEEE Trans. Antennas Propag.}, vol.~73, no.~8, pp.
  5945--5960, Aug. 2025.

\bibitem{ding2025PA}
Z.~Ding, R.~Schober, and H.~Vincent~Poor, ``Flexible-antenna systems: A
  pinching-antenna perspective,'' \emph{IEEE Trans. Commun.}, vol.~73, no.~10,
  pp. 9236--9253, Oct. 2025.

\bibitem{ding2026PA}
Y.~Liu, Z.~Wang, X.~Mu, C.~Ouyang, X.~Xu, and Z.~Ding, ``Pinching-antenna
  systems: Architecture designs, opportunities, and outlook,'' \emph{IEEE
  Commun. Mag.}, vol.~64, no.~1, pp. 190--196, Jan. 2026.

\bibitem{xu2025PA}
Y.~Xu, Z.~Ding, and G.~K. Karagiannidis, ``Rate maximization for downlink
  pinching-antenna systems,'' \emph{IEEE Wireless Commun. Lett.}, vol.~14,
  no.~5, pp. 1431--1435, May 2025.

\bibitem{zheng2026RA}
B.~Zheng, Q.~Wu, T.~Ma, and R.~Zhang, ``Rotatable antenna-enabled wireless
  communication: Modeling and optimization,'' \emph{IEEE Trans. Commun.},
  vol.~74, pp. 6825--6842, 2026.

\bibitem{zheng2026RACS}
B.~Zheng, T.~Ma, C.~You, J.~Tang, R.~Schober, and R.~Zhang, ``Rotatable antenna
  enabled wireless communication and sensing: Opportunities and challenges,''
  \emph{IEEE Wireless Commun.}, vol.~33, no.~3, pp. 134--141, Jun. 2026.

\bibitem{zhou2025RA}
C.~Zhou, C.~You, B.~Zheng, X.~Shao, and R.~Zhang, ``Rotatable antennas for
  integrated sensing and communications,'' \emph{IEEE Wireless Commun. Lett.},
  vol.~14, no.~9, pp. 2838--2842, Sep. 2025.

\bibitem{MA2024zhu}
L.~Zhu, W.~Ma, B.~Ning, and R.~Zhang, ``Movable-antenna enhanced multiuser
  communication via antenna position optimization,'' \emph{IEEE Trans. Wireless
  Commun.}, vol.~23, no.~7, pp. 7214--7229, Jul. 2024.

\bibitem{li2026MA}
Z.~Xiao, Z.~Li, L.~Zhu, B.~Ning, D.~Benevides~da Costa, X.-G. Xia, and
  R.~Zhang, ``Movable antenna aided {NOMA}: Joint antenna positioning,
  precoding, and decoding design,'' \emph{IEEE Trans. Wireless Commun.},
  vol.~25, pp. 4595--4612, 2026.

\bibitem{wu2025move}
Y.~Wu, D.~Xu, D.~Wing Kwan~Ng, W.~Gerstacker, and R.~Schober, ``Globally
  optimal movable antenna-enabled multiuser communication: Discrete antenna
  positioning, power consumption, and imperfect {CSI},'' \emph{IEEE Trans.
  Commun.}, vol.~73, no.~10, pp. 9903--9923, Oct. 2025.

\bibitem{pi2024mu}
Z.~Xiao, X.~Pi, L.~Zhu, X.-G. Xia, and R.~Zhang, ``Multiuser communications
  with movable-antenna base station: Joint antenna positioning, receive
  combining, and power control,'' \emph{IEEE Trans. Wireless Commun.}, vol.~23,
  no.~12, pp. 19\,744--19\,759, Dec. 2024.

\bibitem{zhang2025mu}
Y.~Zhang, Y.~Zhang, L.~Zhu, S.~Xiao, W.~Tang, Y.~C. Eldar, and R.~Zhang,
  ``Movable antenna-aided hybrid beamforming for multi-user communications,''
  \emph{IEEE Trans. Veh. Technol.}, vol.~74, no.~6, pp. 9899--9903, Jun. 2025.

\bibitem{mao2025MA}
H.~Mao, X.~Pi, L.~Zhu, Z.~Xiao, X.-G. Xia, and R.~Zhang, ``Sum rate
  maximization for movable antenna enhanced multiuser covert communications,''
  \emph{IEEE Wireless Commun. Lett.}, vol.~14, no.~3, pp. 611--615, March 2025.

\bibitem{cao2024MA}
Z.~Xiao, S.~Cao, L.~Zhu, Y.~Liu, B.~Ning, X.-G. Xia, and R.~Zhang, ``Channel
  estimation for movable antenna communication systems: A framework based on
  compressed sensing,'' \emph{IEEE Trans. Wireless Commun.}, vol.~23, no.~9,
  pp. 11\,814--11\,830, Sep. 2024.

\bibitem{cao2025MA}
Z.~Xiao, S.~Cao, L.~Zhu, B.~Ning, X.-G. Xia, and R.~Zhang, ``Channel estimation
  for movable antenna aided wideband communication systems based on compressed
  sensing,'' \emph{IEEE Trans. Wireless Commun.}, vol.~25, pp.
  16\,964--16\,981, 2026.

\bibitem{ma2023MA}
W.~Ma, L.~Zhu, and R.~Zhang, ``Compressed sensing based channel estimation for
  movable antenna communications,'' \emph{IEEE Commun. Lett.}, vol.~27, no.~10,
  pp. 2747--2751, Oct. 2023.

\bibitem{han2025re}
S.~Han, Z.~Zhou, L.~Zhu, J.~Ding, X.~Yang, B.~Jiao, and R.~Zhang, ``Movable
  antenna-aided two-way relay system,'' \emph{IEEE Wireless Commun. Lett.},
  vol.~14, no.~12, pp. 4167--4171, Dec. 2025.

\bibitem{li2025re}
N.~Li, W.~Mei, P.~Wu, B.~Ning, and L.~Zhu, ``Movable antenna enhanced {DF} and
  {AF} relaying systems: Performance analysis and optimization,'' \emph{IEEE
  Trans. Commun.}, vol.~73, no.~12, pp. 13\,239--13\,255, Dec. 2025.

\bibitem{zhou2025re}
W.~Zhou, D.~Yang, Y.~Xu, L.~Xiao, F.~Wu, and T.~Zhang, ``Movable antenna array
  for improving {AAV} relaying networks,'' \emph{IEEE Wireless Commun. Lett.},
  vol.~14, no.~12, pp. 4127--4131, Dec. 2025.

\bibitem{li2024re}
N.~Li, W.~Mei, B.~Ning, and P.~Wu, ``Movable antenna enhanced {AF} relaying:
  Two-stage antenna position optimization,'' in \emph{2024 IEEE Globecom
  Workshops (GC Wkshps)}, Dec. 2024, pp. 1--6.

\bibitem{zhu2024MA}
L.~Zhu, W.~Ma, and R.~Zhang, ``Modeling and performance analysis for movable
  antenna enabled wireless communications,'' \emph{IEEE Trans. Wireless
  Commun.}, vol.~23, no.~6, pp. 6234--6250, Jun. 2024.

\bibitem{wi2011DF}
H.~Wicaksana, S.~H. Ting, Y.~L. Guan, and X.-G. Xia, ``Decode-and-forward
  two-path half-duplex relaying: Diversity-multiplexing tradeoff analysis,''
  \emph{IEEE Trans. Commun.}, vol.~59, no.~7, pp. 1985--1994, Jul. 2011.

\bibitem{yang2026ee}
X.~Yang, S.~Gong, S.~Xie, H.~Liu, T.~Yu, and C.~Xing, ``Energy efficiency
  optimization for {MA}-enabled hybrid {MIMO} communication networks,''
  \emph{IEEE Internet Things J.}, vol.~13, no.~1, pp. 1054--1069, Jan. 2026.

\bibitem{fuji2018td3}
S.~Fujimoto, H.~Hoof, and D.~Meger, ``Addressing function approximation error
  in actor-critic methods,'' in \emph{Proc. Int. Conf. Mach. Learn.}, 2018, pp.
  1587--1596.

\bibitem{han2019drl}
S.~Han and Y.~Sung, ``Dimension-wise importance sampling weight clipping for
  sample-efficient reinforcement learning,'' in \emph{Proc. Int. Conf. Mach.
  Learn.}, 2019, pp. 2586--2595.

\bibitem{math}
E.~Björnson, M.~Bengtsson, and B.~Ottersten, ``Optimal multiuser transmit
  beamforming: A difficult problem with a simple solution structure,''
  \emph{IEEE Signal Process. Mag.}, vol.~31, no.~4, pp. 142--148, Jul. 2014.

\end{thebibliography}

\end{document}